\documentclass[iop, apj,numberedappendix]{emulateapj}

\newcommand\submitms{n}		
\newcommand\bibinc{y}		

\usepackage{wasysym}
\usepackage{subeqnarray}

\newcommand{\ie}{i.e.\ }
\newcommand{\eg}{e.g.\ }
\newcommand{\p}{\partial}

\newcommand{\brak}[1]{\langle #1\rangle}

\newcommand{\gsc}{\;\mathrm{g/cm^{2}}}

\newcommand{\au}{R_{\rm AU}}

\newcommand{\ts}{t_{\rm stop}}
\newcommand{\ls}{\ell_{\rm stop}}
\newcommand{\te}{t_{\rm eddy}}
\newcommand{\ve}{V_{\rm eddy}}
\newcommand{\led}{\ell_{\rm eddy}}
\newcommand{\taus}{\tau_{\rm s}} 

\newcommand{\vdr}{v_{\rm drift}}
\newcommand{\tdr}{t_{\rm drift}}

\newcommand{\Sc}{{\rm Sc}}
\newcommand{\gs}{_{\rm g}}
\newcommand{\ps}{_{\rm p}}

\newcommand{\mk}{\mathcal{K}}
\newcommand{\mF}{\mathcal{F}_{\rm W}}
\newcommand{\Fg}{F}
\newcommand{\QT}{Q_{\rm T}}
\newcommand{\QR}{Q_{\rm R}}
\newcommand{\QDp}{Q_D}
\newcommand{\QDg}{Q_{\alpha}}

\newcommand{\fgm}{_{\rm fgm}}
\newcommand{\fdr}{f_{\rm in}}

\newcommand{\vc}[1]{\mbox{\boldmath{$#1$}}}

\DeclareMathSymbol{\varOmega}{\mathord}{letters}{"0A}
\DeclareMathSymbol{\varSigma}{\mathord}{letters}{"06}
\DeclareMathSymbol{\varPsi}{\mathord}{letters}{"09}

\newcommand{\Eq}[1]{equation\,(\ref{#1})}
\newcommand{\Eqs}[2]{equations (\ref{#1}) and~(\ref{#2})}
\newcommand{\Eqss}[2]{equations (\ref{#1})--(\ref{#2})}

\newcommand{\Fig}[1]{Fig.~\ref{#1}}
\newcommand{\Figs}[2]{Figs.~\ref{#1} and \ref{#2}}
 
\newcommand{\Tab}[1]{Table \ref{#1}}

\slugcomment{ApJ, accepted}

\shorttitle{Planetesimals via Secular GI}
\shortauthors{Youdin}

\begin{document}

\title{On the Formation of Planetesimals via Secular Gravitational Instabilities with Turbulent Stirring}
\author{Andrew N.\ Youdin}
\affil{Harvard Smithsonian Center for Astrophysics, 60 Garden Street, Cambridge, MA 02138}

\email{ayoudin@cfa.harvard.edu}

\begin{abstract}
We study the gravitational instability (GI) of small solids in a gas disk as a mechanism to form planetesimals.   Dissipation from gas drag introduces secular GI, which proceeds even when standard GI criteria for a critical density or Toomre's $Q$ predict stability.   We include the stabilizing effects of turbulent diffusion, which suppresses small scale GI.   The radially wide rings that do collapse contain up to $\sim 0.1$ Earth masses of solids.  Subsequent fragmentation of the ring (not modeled here) would produce a clan of chemically homogenous planetesimals.
Particle radial drift time scales (and, to a lesser extent, disk lifetimes and sizes) restrict the viability of secular GI to disks with weak turbulent diffusion, characterized by $\alpha \lesssim 10^{-4}$.   Thus midplane dead zones are a preferred environment.
Large solids with radii $\gtrsim 10$ cm collapse most rapidly because they partially decouple from the gas disk.  Smaller solids, even below $\sim$ mm-sizes could collapse if particle-driven turbulence is weakened by either localized pressure maxima or  super-Solar metallicity.  Comparison with simulations that include particle clumping by the streaming instability shows that our linear model underpredicts rapid, small scale gravitational collapse.  Thus the inclusion of more detailed gas dynamics promotes the formation of planetesimals.  We discuss relevant constraints from Solar System and accretion disk observations.

\end{abstract}

\keywords{hydrodynamics --- instabilities --- planetary systems: formation
--- planetary systems: protoplanetary disks --- solar system: formation --- turbulence}

\section{Introduction}
The discovery of a diverse population of extrasolar planets\footnote{See http://exoplanet.eu, compiled by Jean Schneider, for an updated catalog with references.} motivates the development of planet formation theories that are (at least) physically consistent and (at best) predictive of the rapidly growing data set.  Forming planetesimals above kilometer sizes is both a crucial first step and testable in our own Solar system.   \citet[hereafter CY10]{cy10} and \citet{houches10} review the main physical processes and theoretical puzzles.

A promising route to planetesimal formation is the clumping of solids by the streaming instability (SI) in protoplanetary disks  \citep{YG05, yj07, jy07}, which then induces gravitational collapse \citep{nature07,jym09}.  SI is a drag instability \citep{gp00} that arises from the aerodynamic back reaction of solids on gas, and does not require self-gravity. 

\begin{figure}[tb] 
\if\submitms y
 \plotone{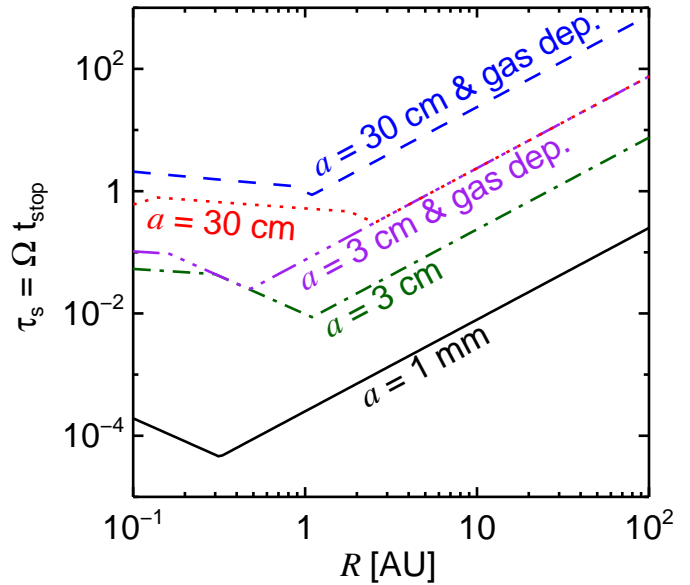}
\else
 \hspace{-1cm}
  \includegraphics[width=3.9in]{f1.eps} 
\fi   
   \caption{Aerodynamic stopping times (normalized by the orbital frequency) vs.\ radial distance in the disk midplane.  For our reference minimum mass model ($\Fg = 1$), we plot $\taus$ for particle sizes: $a = 1$ mm (\emph{black solid curve}), $a = 3$ cm (\emph{green dot-dashed curve}) and $a = 30 cm$ (\emph{red dotted curve}).  For a gas depleted disk ($\Fg = 0.1$) we again plot $\taus$ for $a = 3$ cm (\emph{purple tripe dot-dashed curve}) and $a = 30$  cm (\emph{blue dashed curve}).  The kinks in the curves correspond to transitions between drag laws, as described in \S\ref{sec:drag} and appendix \ref{sec:dragapp}.}
   \label{fig:ts}
\end{figure}

Strong SI-induced clumping appears to require fairly large solids with $\taus \sim 1$, where
\begin{equation} 
\taus \equiv \varOmega \ts
\end{equation} 
measures the dynamical relevance of drag forces with $\ts$ the time scale for damping particle relative motion and $\varOmega$ the Keplerian orbital frequency.  Thus $\taus$ increases with particle radius $a$.  In standard disk models, marginal coupling, $\taus = 1$, corresponds to  $a \lesssim 1$ meter near 1 AU, and smaller sizes in the outer disk where gas densities are lower (see \Fig{fig:ts}).  For small solids with $\taus \ll 1$, SI operates but it stirs particles without producing strong clumping \citep{jy07, bs10a}.  Most other particle concentration mechanisms also become inefficient for $\taus \ll 1$ (see CY10, \citealp{houches10}).

It is desirable to find a planetesimal formation mechanism that operates when $\taus \ll 1$.  Protoplanetary disks initially contain sub-micron-sized grains.  Collisional sticking alone appears unlikely to produce meter-sized bodies \citep{benz00,ys02,st03,you04,zsom10}.  Furthermore, many primitive meteorites are composed primarily of chondrules, inclusions with $a \sim 0.1$ --- $1$ mm \citep{russ06}.  \citet{chs08} propose that turbulent eddies concentrate chondrules into clumps that become 10 --- 100 km planetesimals.  As discussed more extensively in CY10, this concentration mechanism normally applies to much smaller scales.  More study of mechanisms to concentrate $\taus \ll 1$ solids is needed.

This work considers gravitational instabilities (GI) that are secular, meaning mediated by dissipation in the form of gas drag.  The original proposals that planetesimals formed by GI \citep{saf69, gw73} applied the standard, dissipationless GI criterion of \citet{too64}.  This approach gives the classic result of $\sim$ kilometer-sized planetesimals.  The fact that gas drag affects GI should not be surprising, especially for $\taus \ll 1$. Section \ref{sec:prev} summarizes previous work on secular GI .

This paper contains an improved treatment of gas turbulence, notably the stabilizing effects of turbulent diffusion.  We use the turbulent stirring model of \citet{yl07}, which includes the effects of orbital oscillations.  Our simplified treatment of gas dynamics does not capture SI or other non-gravitational concentration mechanisms.  However, we do vary the disk ``metallicity"
\begin{equation}\label{eq:Z} 
Z \equiv \Sigma/\Sigma\gs \, ,
\end{equation}
the ratio of particle (unsubscripted) to gas surface densities.  Large $Z$ values mimic particle concentration or gas depletion on large scales (compared to the wavelength) and accelerate secular GI.

We emphasize one general result from our analysis.  The critical or Roche-like\footnote{The actual Roche density refers to tidal disruption of a sphere and is numerically larger (see CY10).  The numerical coefficient in \Eq{eq:rhoR} is from the buckling modes of \citet{sek83}, which we discuss further in \S\ref{sec:prev}.} density
\begin{equation}\label{eq:rhoR}
\rho_{\rm R} \simeq 0.6 {M_\ast \over R^3} 
\end{equation} 
is \emph{not} a useful discriminant for gravitational collapse when drag forces are significant, as they are for planetesimal formation.  We use $M_\ast$ for the central stellar mass and $R$ for the radial distance from that star.  Dissipation allows collapse to proceed for particle densities $\rho \ll \rho_{\rm R}$.  We prove this result in \S\ref{sec:proof} and further demonstrate it in \Fig{fig:almaxrho}.  Moreover, if $\rho > \rho_{\rm R}$ in only a small volume, then collapse is avoidable.  Gas drag limits collapse speeds to the terminal velocity, and turbulent diffusion smoothes small scale density perturbations.  

Since secular GI can require many orbits, it is restricted by radial drift speeds and ultimately disk lifetimes.   Observations find protoplanetary disks with ages up to several million years  \citep{hern07b}.  Planetesimal formation within 1 Myr should still allow time for the growth of gas and ice giants by core accretion \citep{ls07}.  The existence of primitive meteoritic inclusions with age spreads of several Myr \citep{russ06} supports a longer time scale for planetesimal formation in the inner Solar System.  

This article is organized as follows.  We start with an order of magnitude derivation of secular GI in \S\ref{sec:OOM}.  We present our full model in \S\ref{sec:form} including: a derivation of the dispersion relation in \S\ref{sec:eqns}, the turbulent stirring model in \S\ref{sec:turbmodel} and the protoplanetary disk model in \S\ref{sec:disk}.  Section \ref{sec:results} gives the main numerical results covering: the variation of GI properties with disk parameters in \S\ref{sec:fixalpha} and a calculation of the strongest levels of turbulence that allow collapse in \S\ref{sec:alphamax}.   Section \ref{sec:GenBehavior} explores the analytic behavior of secular secular GI and the transition to standard GI.   We discuss several aspects of our results in \S\ref{sec:disc} including a comparison to simulations in \S\ref{sec:sims},  Solar System constraints in \S\ref{sec:SS} and metallicity thresholds in \S\ref{sec:growthdisc}.  Section \ref{sec:conc} concludes with a brief summary.  Appendix \ref{sec:dragapp} addresses turbulent drag laws and the behavior of secular GI in less relevant regions of parameter space is relegated to Appendix \ref{sec:regapp}.

\section{Secular Collapse of Small Solids}\label{sec:OOM}
We begin with an order-of-magnitude derivation of the main result of this paper: the gravitational collapse of small solids in a gas disk.   Small means that drag forces are strong with $\taus \ll 1$.
Though probably less relevant for planetesimal formation, secular GI also exists for $\taus \gg 1$, see  \S\ref{sec:regII}.

Consider a ring-like surface density perturbation of amplitude $\sigma$ and radial length scale $\lambda$.  The radial gravity at the edge of the perturbation is $g_R \sim G \sigma$.  If this result is unfamiliar, recall that the acceleration towards an infinitely long wire of linear mass density $\Lambda$ is $2 G \Lambda/\lambda$ at a distance $\lambda$.  Then identify $\sigma \sim \Lambda/\lambda$ as if the ring were concentrated in a thin wire, which  can be approximated as straight and infinitely long because $\lambda \ll R$, the radial distance to the star.

The particles' response to self-gravity is mediated by gas drag.  Radial sedimentation to the center of the ring occurs at the terminal velocity
\begin{equation} \label{eq:driftapp}
u = g_R \ts \sim G \sigma \ts \, .
\end{equation}
It is reasonable to ask why the radial gravity induces radial collapse instead of merely shifting the orbital speed by adjusting centripetal balance.  The reason is that the drag on this extra orbital motion would would give radial infall that exceeds the terminal velocity limit, hence \Eq{eq:driftapp} applies.  This explanation 
assumes the pressure-supported gas remains stationary in response to the particle perturbations, a critical issue that we discuss further in \S\ref{sec:prev}.
 
By mass continuity, the amplitude of the density perturbation grows at a rate
\begin{equation} \label{eq:sapprox}
s \sim { \varSigma u \over \sigma \lambda} \sim {G \varSigma \ts \over \lambda}\, .
\end{equation} 
To understand whether and how fast collapse occurs, we must consider stabilizing influences that restrict the allowable $\lambda$.   The excess angular momentum in a perturbation is removed very rapidly, in $\ts \ll 1/s$ as we can verify shortly.  Thus there is no obstacle to long wavelength collapse, and instability is guaranteed.   We now consider impediments to small scale collapse that set the minimum $\lambda$.

\subsection{Effective Pressure Support: Less Relevant}
We first consider the stabilization by the effective pressure of particle random velocities, $c$.  Gravity, $g_R$ exceeds the effective pressure acceleration, ${c^2 \sigma /( \varSigma \lambda)}$,
for wavelengths above
\begin{equation} 
\lambda_P \sim {c^2 \over G \varSigma}  \, .
\end{equation} 

From \Eq{eq:sapprox} the fastest growth rate allowed by effective pressure support is
\begin{equation} 
s_{P} \sim \left(G \varSigma \over c\right)^2 \ts \sim \varOmega {\taus \over \QT^2}\, .
\end{equation} 
$\QT = c \varOmega/(\pi G \varSigma)$ is the standard \citet{too64} gravitational stability parameter, but unlike standard GI we have no requirement that $\QT < 1$.

\subsection{Diffusive Support: Dominant}
Next we consider the stabilizing effect of particle diffusivity, $D$, due to stirring by gas turbulence. We express
\begin{equation} 
D \approx c^2 /\varOmega\, .
\end{equation} 
as the product of the particle speed, $c$, and path length, $c/\varOmega$, which arises because disk turbulence randomizes the particle trajectory in an eddy turnover time  $\te = 1/\varOmega$.  See \S\ref{sec:turbmodel} for more details on turbulent stirring.   

For the collapse rate of \Eq{eq:sapprox} to outpace the diffusive spreading time scale $\lambda^2 /D$, wavelengths must exceed
\begin{equation} \label{eq:lDapprox}
\lambda_D \sim {D \over G \varSigma \ts} \sim {\lambda_P \over \taus}\, .
\end{equation} 
Thus $\lambda_D \gg \lambda_P$, by our assumption of strong drag.  Diffusion is the more effective stabilizing influence that sets the actual minimum $\lambda$.

Secular GI thus has a collapse rate
\begin{equation} \label{eq:sDapprox}
s \sim {(G \varSigma \ts)^2 \over D} \sim \varOmega \left(\taus \over \QT \right)^2\, , 
\end{equation} 
that is slower than $s_P$ by $\taus \ll 1$.   Growth becomes significantly slower as particle size, and thus $\taus$, declines.

Previous studies of dissipative collapse that neglected turbulent diffusion \citep{war76, war00,y05a,y05b} were thus overly optimistic.  However the qualitative result that collapse does not require $\QT < 1$ remains.

Armed with this basic description of the instability, some readers may wish to skip ahead to the numerical results of \S\ref{sec:results} and the figures.   To better understand the model ingredient and caveats, we do recommend coming back to \S\ref{sec:form}.

\section{Formulation of the Problem}\label{sec:form}
\subsection{Dynamical Equations}\label{sec:eqns}
We consider the local, linear dynamics of solid particles in a protoplanetary disk subject to gas drag, turbulent mixing and self-gravity.  We adopt the local shearing box model for a patch of the disk \citep{glb65II} with a frame centered on a fiducial radius that rotates with the local Kepler frequency $\varOmega$.  The local radial and azimuthal coordinates are $x$ and $y$, respectively.  We decompose the particle surface density $\Sigma\ps = \Sigma +\sigma$ and velocity fields ${\bf V} = {\bf V}_o + u\hat{x} + v\hat{y}$ into time-steady backgrounds --- constant $\Sigma$ and the Keplerian shear ${\bf V}_o = -(3/2) \varOmega x \hat{y}$ ---  plus small amplitude perturbations.  The perturbations are axisymmetric as appropriate for slow motions stretched by shear.  Evolution in time, $t$, follows the linearized, height-integrated equations of continuity and force:
\begin{subeqnarray} \label{eq:eom}
 \frac{\p \sigma}{\p t} + \Sigma \frac{\p u}{\p x}  &=& D \frac{\p^2 \sigma}{\p x^2}\, , \slabel{eq:cont}\\
\frac{\p u}{\p t} - 2\varOmega v  &=&   - \frac{u}{\ts }  - {c^2 \over \Sigma} \frac{\p \sigma}{\p x} -\frac{\p \phi}{\p x} \, , \slabel{eq:xmom}\\ 
\frac{\p v}{\p t} + \varOmega u/2  &=& - \frac{v}{\ts } \slabel{eq:ymom}\, ,  
\end{subeqnarray}
The right hand side of \Eq{eq:cont} describes mass diffusion, $D$, induced by turbulent gas.  The effects of this term have also been considered by Shariff \& Cuzzi (personal communication).   The first term on the RHS of \Eqs{eq:xmom}{eq:ymom} are drag forces on particle motion relative to the static gas background.  The aerodynamic stopping time, $\ts$ is evaluated in terms of particles sizes and gas densities in \S\ref{sec:drag}.
The last two terms on the RHS of \Eq{eq:xmom} are standard in analyses of gravitational instability (GI): the effective pressure acceleration from (isothermal) random particle velocities, $c$, and the acceleration from the disk's self-gravitational potential $\phi$.  The validity of these model equations is discussed in \S\ref{sec:valid}.

We give the solution to Poisson's equation in terms of Fourier amplitudes which are denoted by tildes and have a spatiotemporal dependence  $\propto \exp(s t + \imath k x) $:
\begin{equation}\label{eq:poiss}
 \tilde{\phi} = - {2 \pi G \tilde{\sigma} \over k} \mathcal{T}(k h)\, ,
\end{equation}
The softening term, $\mathcal{T}(k h) = 1/(1+k h)$ mimics the effects of finite particle layer thickness, $h$ \citep{shu84}.  This factor connects the thin disk potential of long waves, $k h \ll 1$, to the three dimensional solution for $k h \gg 1$.  While the modes studied in this paper are generally long, $\mathcal{T}$ insures against the appearance of artificially short waves in a height-integrated model.

\subsubsection{Dimensionless Dispersion Relation}
We use the standard GI wavelength
\begin{equation} \label{eq:lg}
\lambda_G \equiv 2 \pi \ell_G \equiv  2 \pi^2 G \varSigma/\varOmega^2
\end{equation}
and the orbital time, $1/\varOmega$, to define dimensionless velocities, $\{U,V\} \equiv  \{u,v\}/(\ell_G\varOmega)$, radial position $X \equiv  x/\ell_G$ and time $T = \varOmega t$.   The growth rate, $s$, and wavenumber, $k = 2 \pi /\lambda$, are non-dimensionalized as
\begin{subeqnarray} 
\mk &\equiv&  k \cdot \ell_G = \lambda_G/\lambda \, , \\ 
\gamma &=& s/\varOmega\, ,
\end{subeqnarray}
so the Fourier dependance becomes: $\exp( \imath \mk X + \gamma T)$.  In terms of dimensionless Fourier amplitudes, the equations of motion read
\begin{subeqnarray} \label{eq:dimless}
(\gamma + \mk^2 \QDp^2) \tilde{\sigma}/\varSigma + i \mk \tilde{U} &=& 0 \, , \label{eq:contdimless}\\
(\gamma + 1/\taus ) \tilde{U} -2\tilde{V} &=& \imath{1-\mF \over \mk}{\tilde{\sigma} \over \varSigma} \, , \slabel{eq:xdimless}\\
(\gamma + 1/\taus) \tilde{V} +\tilde{U}/2 &=& 0\, . \slabel{eq:ydimless}
\end{subeqnarray}
The cubic dispersion relation for the (possibly complex) growth rate $\gamma(\mk)$ is
\begin{equation} \label{eq:disp}
\left[\left(\gamma + \taus^{-1}\right)^2 + 1\right] \left(\gamma + \mk^2\QDp^2 \right) = \left(\gamma + \taus^{-1}\right) \left(1-\mF\right)\, . 
\end{equation}  

We follow \citet{war76,war00} in parameterizing the strength of self-gravity with the function
\begin{subeqnarray} \label{eq:F}
\mF &\equiv&  \frac{\varOmega^2 - 2\pi G\varSigma k \mathcal{T}(kh) + k^2c^2}{\varOmega^2} \\
 &=& 1 - {2 \mathcal{K} \over 1 + \mathcal{K}\QR} + \QT^2 \mathcal{K}^2 \, .
 \end{subeqnarray}
Smaller $\mF$ indicates stronger self gravity.  

At this point, the model requires four input parameters.  These are the stopping time $\taus$ plus three gravitational stability parameters;
\begin{subeqnarray}\label{eq:Qgen}
\QT &\equiv& {c \varOmega \over \pi G \varSigma}\, , \slabel{eq:QTg}\\ 
\QR &\equiv& {h \varOmega ^2 \over \pi G \varSigma} \approx 0.2 {\rho_{\rm R} \over \rho}\, , \slabel{eq:QRg}\\
\QDp  &\equiv& {\sqrt{D}\varOmega^{3/2} \over \pi G \varSigma} \, ;
\end{subeqnarray}
 which incorporate the effective pressure (with T for Toomre), layer thickness (with R for Roche) and turbulent diffusion of solids (D), respectively.  We express $\QR$ in terms of the critical density of \Eq{eq:rhoR} and the midplane particle density $\rho = \Sigma/(\sqrt{2\pi} h)$.
    
The three $Q$ values are not independent.  We relate them to the level of  turbulent stirring in \S\ref{sec:turbmodel}, which has the added benefit of reducing the number of input parameters to two.  This approach allows the complete survey of solutions in \S\ref{sec:GenBehavior}.  First we describe some general properties of the dispersion relation.

\subsubsection{Standard GI: No Drag or Diffusion}\label{sec:standardGI}
We recover standard axisymmetric GI by ignoring gas drag, $\taus \rightarrow \infty$,  and turbulent diffusion, $\QDp = 0$.  The dispersion relation takes the simple form
\begin{equation} \label{eq:GIdr}
\gamma(\gamma^2 + \mF) = 0\, ,
\end{equation} 
giving a neutral mode (NM) with $\gamma = 0$ and a pair of density waves (DW) with $\gamma = \pm\sqrt{-\mF}$.  In terms of the dimensional frequency $\omega = \imath s$, the DW solution obeys the familiar dispersion relation:
\begin{equation} 
\omega^2 = \varOmega^2 - 2\pi G \varSigma k \cdot \mathcal{T}(kh) + k^2 c^2
\end{equation} 
The DW solutions correspond to traveling waves if $\mF > 0$, and gravitational collapse for $\mF< 0$.  In the thin disk limit where $\mathcal{T} \rightarrow 1$, we recover the \citet{too64} instability criterion, $\QT <  Q_{\rm T,crit} = 1$, with the marginally unstable wavenumber, $\mk_{\rm crit}= 1$. 
In physical units $\lambda = \lambda_G $.

We now consider the effects of our finite thickness correction.   First note that $\mathcal{T}$ enforces a Roche-like limit.  Even for a perfectly cold layer with $\QT = 0$, instability ($\mF < 0$) requires $\QR \le 2$. 
If we fix $\QT = \QR$ (the standard  $c = \varOmega h$ relation from Keplerian orbital oscillations or hydrostatic balance) then  $Q_{T, \rm crit} \approx 0.55$ and $\mathcal{K}_{\rm crit} \approx 1.2$.  This solution ignores the compression of the layer by self-gravity which enhances the vertical gravity by a factor
 \begin{equation} \label{eq:psi}
\psi \equiv 1 + {2 \pi G \Sigma \over \varOmega^2 h} =  1 + {2 \over \QR} \, .
\end{equation} 
Vertical hydrostatic balance then gives a relation $\QT = \QR \sqrt{1+2/\QR}$ (as in \citealp{war00}).  In this case $Q_{\rm T, crit} \approx 0.77$ with  $Q_{\rm R, crit} \approx 0.26$ and $\mathcal{K}_{\rm crit} \approx 1.05$, in good agreement with more rigorous analyses \citep{glb65II}. Thus the finite thickness correction to Toomre instability is relatively modest when vertical compression is included.  Secular GI has much longer wavelengths, making the finite thickness corrections insignificant in most cases of interest.  

\subsubsection{Proof of Generic Instability}\label{sec:proof}
We now prove that the full dispersion relation of \Eq{eq:disp} always yields a growing mode for some wavelengths, for \emph{any} choice of stability parameters or stopping time.  This result is fundamentally different from  standard GI, as we just reviewed.  The existence of a growing mode does not assure its  astrophysical relevance, so we impose extra constraints in \S\ref{sec:relevant}.  Nevertheless, the following proof demonstrates that instability is quite robust and does not depend, \eg on the specific turbulence model  we adopt in \S\ref{sec:turbmodel}.  Moreover we know that a numerical root finding algorithm has failed if it does not identify a growing mode.

Since the dispersion relation is a cubic in $\gamma$ with real coefficients, we can use Descartes' ``rule of signs" to show that there is always at least one real, positive root (and thus a growing mode).   We write \Eq{eq:disp}  as
\begin{equation} 
\gamma^3 + C_2 \gamma^2 + C_1 \gamma + C_0 =0
\end{equation}
with
\begin{eqnarray*}  
C_2 &=& 2/\taus + \mk^2 \QDp^2  > 0\\
C_1 &=& \mF + \taus^{-2} + 2 \mk^2 \QDp^2/\taus \\
C_0 &=& (\mF-1)/\taus + C'\mk^2/\taus \\
C' &=&  \QDp^2 (\taus + \taus^{-1})>0
\end{eqnarray*} 
We use the long wave limit where $\mF  \rightarrow 1-2\mk$ as $\mk \rightarrow 0$.
We have $C_2 > 0$ and $C_1 > 0$ since $\mF>0$ for long waves.  The $C_0$ coefficient is negative if $ C' \mk^2  < 1 - \mF$, which holds because $ C'\mk^2< 2\mk$ as $\mk \rightarrow 0$.  Thus we can ensure (for some $\mk$) one, and only one, sign change in the coefficients, which guarantees a positive root.  Q.E.D.

\subsection{Turbulence Model}\label{sec:turbmodel}
We will now describe the response of solid particles to turbulent stirring.  This section culminates in \Eq{eq:Qs}, which expresses the self-gravity parameters $\QT$, $\QR$ and $\QDp$ in terms of the stopping time $\taus$ and a parameter
\begin{eqnarray} \label{eq:QDg}
\QDg &=& {\sqrt{D\gs}\varOmega^{3/2} \over \pi G \varSigma} = \sqrt{\alpha} {c\gs \varOmega \over \pi G \varSigma} \,  
\end{eqnarray}  
that compares turbulent stirring to particle self-gravity.  Here $D\gs$ is the strength of turbulent diffusion in the gas, and 
\begin{equation}\label{eq:alpha}
\alpha = D\gs \varOmega/c\gs^2 
\end{equation} 
is a more familiar dimensionless measure of turbulent intensity with $c\gs$ the sound speed in the gas.  We  discuss in \S\ref{sec:conc} why this  diffusive $\alpha$ can be much smaller than the (purposefully similar, but not equivalent) parameter used in studies of accretion disks. 

This subsection makes heavy use of the results of  \citet[hereafter YL07]{yl07}.  We extend their results in our \Eqss{eq:hpsi}{eq:hfull} by including self-gravity in the scale height calculation.

\subsubsection{Turbulence and $\alpha$ Values}\label{sec:turbgas}
We adopt a relatively simple model for disk turbulence.  As in Kolmogorov turbulence, the largest eddies dominate the energy budget.  We further assume that $\te = 1/\varOmega$, \ie the turnover time of these integral scale eddies equals the orbital time scale.  This assumption requires a turbulent speed $\ve = \sqrt{\alpha} c\gs$ and eddy length $\led = \sqrt{\alpha} c\gs/\varOmega$ so that $\te = \led/\ve$ and the gas diffusivity $D\gs = \ve \led$ satisfies \Eq{eq:alpha}, as noted by \cite{cuz01}.  

We further assume that the turbulent kinetic energy is isotropic with partial correlation between the radial, $u\gs$, and azimuthal, $v\gs$, turbulent speeds.
Specifically, we take
\begin{equation} \label{eq:turbmodel}
\brak{u\gs^2} = \brak{v\gs^2} = \brak{w\gs^2} = 4 \brak{u\gs v\gs}  \equiv \ve \equiv \alpha c\gs^2
\end{equation} 
where $w\gs$ is the vertical turbulent speed and angular brackets indicate time-averages.
While the velocity correlations are of modest significance for particle stirring (see YL07) they are a primary concern for the angular momentum transport and the accretion of gas onto the protostar, see \S\ref{sec:conc}.

\subsubsection{Particle Response to Turbulence}\label{sec:turbresponse}
The radial  particle diffusion coefficient is given by equation (36) of YL07 .  With our \Eq{eq:turbmodel} 
\begin{equation}\label{eq:D}
D = {1 + \taus + 4 \taus^2 \over (1+\taus^2)^2}D\gs\, .
\end{equation} 
The detailed $\taus$ dependence will vary with the exact properties of the turbulent cascade.  The important result is that $\taus \ll 1$ solids diffuse at a rate similar to any passive contaminant, $D \simeq D\gs$.  Particle clumping in turbulence could lower $D$, but we do not have a good model for this effect.  As particles decouple and $\taus \gg 1$, diffusivity drops as $D/D\gs \propto \taus^{-2}$.  Thus a rough value of the so called Schmidt number is $\Sc \equiv D\gs/D \sim 1 + \taus^2/4$.  This result differs from the often used, but incorrect, value of $\Sc_{\rm CDC} = 1 + \taus$ in \citet{cdc93}, see YL07.

Since we are interested in radial pressure support, the relevant particle dispersion speed, $c$, is the radial component.  We use 
\begin{equation}
c = {\sqrt{1+2\taus^2+(5/4)\taus^3} \over 1+\taus^2}\sqrt{\alpha} c\gs\, . \label{eq:c} 
\end{equation} 
A similar result for $c$ follows from equation (33a) of YL07 and our \Eq{eq:turbmodel}.  In \Eq{eq:c} we adopt a simpler $\taus$ dependence. This choice is purely for convenience and gives indistinguishable numerical results. These details are not crucial.  \citet{vjmr80} showed that the same basic scaling, $c \sim \ve/ (1 + \taus)^{1/2}$, holds when the orbital dynamics included by YL07 are ignored.  

The particle scale height, $h$, usually obeys the well known result
\begin{equation} \label{eq:hstd}
h =  h_{\rm std} \equiv \sqrt{\alpha \over \taus} h\gs\, .
\end{equation} 
where $h\gs = c\gs/\varOmega$ is the hydrostatic gas scale height (for weak self-gravity).  Intuitively the particle layer narrows for weaker turbulence or more loosely coupled solids.  \citet{cdc93} derive the result for $\taus \ll 1$ by comparing settling and diffusive time scales.  \citet{cfp06} show that the same scaling also holds for $\taus \gg 1$ using both simulations and velocity diffusion arguments.  

Corrections to $h_{\rm std}$ arise from self-gravity and the perfect mixing limit.  Including self- gravity  gives 
\begin{equation} \label{eq:hpsi}
h = h_\psi = \sqrt{\alpha \over \taus \psi}h\gs
\end{equation} 
with the amplification of vertical gravity by $\psi$ from \Eq{eq:psi}.  To derive this result (which as far as we know is new) for $\taus \ll 1$, equate the shortened settling time $t_{\rm sett} = 1/(\varOmega \taus \psi)$ with the diffusion time $t_{\rm diff} = h^2/D\gs$.  The possibility that turbulent intensity varies with self-gravity is not considered here, but could be included to model gravito-turbulence.  The same expression holds for $\taus \ll 1$, where the vertical oscillation frequency shortens to $\omega_z = \varOmega/\sqrt{\psi}$.   The random velocity, $c_z = \sqrt{\alpha/\taus}c\gs$, imparted by turbulent fluctuations gives $h_\psi = c_z/\omega_z$, which obeys \Eq{eq:hpsi}.  Since $\psi$ itself depends on $h$, the explicit solution reads
\begin{equation} 
h_\psi = \sqrt{\ell_G^2 + {\alpha \over \taus} h\gs^2} - \ell_G
\end{equation}
which reduces to \Eq{eq:hstd}  for small $\ell_G$, which is defined in \Eq{eq:lg}.

Perfect mixing with the gas imposes an upper limit $h \leq h\gs$.  Following YL07 [their equation (28)] we enforce this limit to give our most general result
\begin{equation} \label{eq:hfull}
h = h_{\rm full} =  \sqrt{\alpha \over \taus \psi + \alpha} h\gs\, .
\end{equation} 
Again there is an implicit $h$ dependence in $\psi$.

 The ratio of particle to gas density, important for drag feedback effects, is
\begin{equation} \label{eq:rhopg}
{\rho \over \rho\gs} = {\Sigma \over \Sigma\gs} {h\gs \over h}  \approx Z \sqrt{\taus \over \alpha} \end{equation}
which uses $h_{\rm std}$ for illustration, though \Eq{eq:hfull} is used in numerical calculations.  

\subsubsection{Dimensionless Stability Parameters}\label{sec:turbQ}
We can now express the $Q$ parameters from \Eq{eq:Qgen} in terms of $\QDg$ and $\taus$.  Applying equations (\ref{eq:D}),  (\ref{eq:c}) and  (\ref{eq:hfull}) gives
\begin{subeqnarray}\label{eq:Qs}
\QT  &=&  {\sqrt{1+2\taus^2+(5/4)\taus^3} \over 1+\taus^2}\QDg \, , \slabel{eq:QT}\\ 
\QR &=& {\sqrt{1+ \QDg^2/\taus(1+\alpha/\taus)}-1 \over {1 + \alpha/\taus}} \, , \slabel{eq:QR}\\
\QDp  &=& {\sqrt{1+\taus+4\taus^2} \over 1 + \taus^2}\QDg\, .
\end{subeqnarray}

Note that \Eq{eq:QR} depends on $\alpha$ in addition to $\QDg$ and $\taus$.  To reduce to only two parameters, we take the $\alpha \ll \taus$ limit, equivalent to using \Eq{eq:hpsi}, and obtain
\begin{equation} 
\QR \approx \sqrt{1 + \QDg^2/\taus} - 1 \, . \label{eq:QRtightapp}
\end{equation} 
This approximation is acceptable on a few counts.  First it is conservative, since we are removing the restriction that $h < h\gs$ and allowing artificially low particle densities.  Second, relevant growth rates typically have $\alpha < \taus$, as seen in \Figs{fig:ts}{fig:almax}.  Finally, even for $\alpha > \taus$, wavelengths are long enough that the finite thickness correction is negligible, $k h = \mk \QR \ll 1$, even with the artificially enhanced $h$.  Thus  \Eq{eq:QRtightapp} is always acceptable for our study, even though \Eq{eq:QR} is more general.

 For $\taus \ll 1$, $\QT \simeq \QDp \simeq \QDg$ is particularly simple.  It reflects the result that tightly coupled particles share the same velocity and diffusion coefficient as the dominant turbulent eddies.   Furtherore $\QR \simeq \QDg/\sqrt{\taus} \gg 1$ for weak self-gravity, reflecting the fact that $h \gg c \varOmega$ for turbulently-stirred small particles.

\subsection{Disk Model}\label{sec:disk}
\subsubsection{A Scalable Minimum Mass Nebula}
For numerical evaluations we adopt a reference minimum mass Solar Nebula model, with scaling factors to explore the effect of changing disk parameters.   In our formulation (similar to CY10 which has further discussion) the surface density of solids and gas and the midplane temperature are
\begin{subeqnarray} 
\varSigma &=& 20  \,Z_\%  F\, \au^{-3/2} \gsc \, , \\
\varSigma\gs &=& 2000\, F\, \au^{-3/2} \gsc \label{eq:Sigmag}\, ,\\
T &=& 200 f_T \au^{-1/2} ~{\rm K}\label{eq:T}\, ,
\end{subeqnarray} 
where $\au \equiv R/{\rm AU}$ is the disk radius in AU.  The  ``metallicity"  $Z \equiv \Sigma/\Sigma\gs \equiv 0.01 Z_\%$ and the mass normalization $F$ are two important adjustable parameters.  Water and methane condensation at ``ice lines" can give sharp transitions in $\Sigma$.  We do not model this in detail, because a range of enrichment processes also affect $Z$.  

We adopt a cool disk temperature characteristic of passive flared disks around Sun-like stars, and fix  $f_T = 1$ in this work.  Our profile is modestly steeper than the $T \propto R^{-3/7}$ of \citep{cg97}, just to give analytic scalings that are both simpler and similar to the classic \citet{hay81} model.  This idealization has little effect, given the weak dependence of our results on such small temperature changes.  

The disk mass within a cutoff radius $R_{\rm out}$ is
\begin{equation} 
M_{\rm disk} = 0.02\, \Fg (1+ Z)  \sqrt{R_{\rm out}/(50 \,{\rm AU})}\, M_\odot .
\end{equation} 
The \emph{gas} disk is gravitationally stable since its Toomre parameter,
\begin{equation} \label{eq:Qg}
Q\gs \equiv {c\gs \varOmega \over \pi G \varSigma\gs} \approx 40 {\sqrt{f_T m_\ast} \over \Fg}\au^{-1/4}\, ,
\end{equation}
satisfies $Q\gs > 1$, with $m_\ast$ the stellar mass in solar masses.   Our measure of particle self-gravity relative to turbulent stirring, from \Eq{eq:QDg},
\begin{equation} \label{eq:QDgnum}
\QDg = \sqrt{\alpha} {Q\gs \over Z} \approx 1.3 \sqrt{\alpha \over 10^{-7}}{\sqrt{f_T m_\ast} \over \Fg Z_\%}\au^{-1/4} \, ,
\end{equation} 
exceeds unity in most disk models.

The midplane gas density is
\begin{equation} 
\rho\gs = {\Sigma\gs \over \sqrt{2 \pi} h\gs} \approx 2 \times10^{-9} \Fg\sqrt{m_\ast \over f_T} \au^{-11/4}~\rm{g \over cm^3}
\end{equation} 
for a vertically isothermal disk, ignoring compression from the weight or self-gravity of the solids \citep{nsh86,ws08}. 

The mean free path for molecular collisions,
\begin{equation} 
\ell\gs \approx 1  \sqrt{f_T \over m_\ast}{1 \over  \Fg} \au^{11/4} ~{\rm cm}\, ,
\end{equation} 
is inversely proportional to $\rho\gs$ and important for the drag forces that we now consider.

\subsubsection{Particle Properties and Gas Drag} \label{sec:drag}
Our model considers a population of particles with uniform radius, $a$, and internal density, $\rho_\bullet = 2~{\rm g/cm}^3$.   The general dependence of drag forces on particle, gas and flow properties is complex, but approximate formulae apply to distinct parameter regimes  \citep{ahn76, stu77}.  The stopping time $\ts \equiv \Delta V/f_{\rm drag}$ is the ratio of the relative flow speed to the drag acceleration.  \Fig{fig:ts}  plots the dimensionless $\taus \equiv \varOmega \ts$ for particle sizes from $a = 1$ mm to $30$ cm for both a minimum mass disk with $\Fg = 1$ and a gas depleted disk with $\Fg = 0.1$.

Epstein drag is the most relevant regime.  It applies for small sizes, $a \lesssim (9/4) \ell\gs$, and low gas densities, with 
\begin{equation} 
\taus = \tau_{\rm Ep}  = {\rho_\bullet a \over \sqrt{2 \pi} \varSigma\gs} \approx 2.5 \times10^{-4} {a_{\rm mm} \over \Fg} \au^{3/2} \, ,
\end{equation}
where $ a_{\rm mm}  = a/{\rm mm}$.
Epstein drag is  ready identifiable in \Fig{fig:ts} as the region where $\taus$ increases with $R$  (outside $0.3$ AU for 1 mm and outside $\sim 2$ AU for 30 cm).

Stokes drag applies to particles too large for Epstein drag, and a Reynolds number, $\mathrm{Re} = a \Delta V/(c\gs \ell\gs) \lesssim 1$.   
The stopping time for viscous Stokes drag
\begin{equation} 
\taus = \tau_{\rm Sto}  = \tau_{\rm Ep}{4 a \over 9 \ell\gs}  \approx 1 \times10^{-3} a_{\rm mm}^2\sqrt{m_\ast \over f_T} \au^{-5/4}\, ,
\end{equation} 
applies in the inner disk, where gas densities are high, and for small solids that do not experience a large headwind.  In \Fig{fig:ts} we identify the Stokes regime for 1 mm and 3 cm solids where $\taus$ declines sharply with radius.  The brief overlap of the $\Fg = 1.0$ and $\Fg = 0.1$ (gas depleted) curves demonstrates that $\tau_{\rm Sto}$ does not depend on gas density.

 In \Fig{fig:ts}, turbulent drag applies in regions where $\taus$ has a relatively flat dependence on $R$.  Appendix \ref{sec:dragapp} discusses non-linear drag regimes.  Unless stated otherwise, the numerical estimates and scalings in this work use Epstein drag.

\subsubsection{Radial Drift}\label{sec:drift}
Particles experience a net inward drift because they encounter a headwind in pressure-supported gas disks.   The orbital speed of a purely gas disk is slower than Keplerian by a speed $\eta \varOmega R$, where
\begin{equation} 
\eta \equiv -{\p P/\p R \over 2 \rho\gs \varOmega^2 R} \approx 1.3 \times10^{-3} {f_T \over m_\ast} \sqrt{\au}
\end{equation} 
measures the radial gradient of the gas pressure, $P$.

The radial drift time scale, defined as the time to reach the star if the speed were to remain constant, is
\begin{equation} \label{eq:tdrift}
\tdr \equiv {R \over \left|\vdr\right|} = \fdr {1 + \taus^2/\fdr \over 2 \eta \taus \varOmega}\, .
\end{equation} 
This equation includes the inertial correction factor \citep{nsh86, yc04}
\begin{equation}\label{eq:fdr} 
\fdr \equiv  \left(  {\rho_{\rm tot} / \rho\gs}\right)^2 = \left( 1 + {\rho / \rho\gs}\right)^2 \, ,
\end{equation} 
which is unity in the test particle limit that, $\rho \ll \rho\gs$.  Test particles with $\taus = 1$ have the shortest drift time
\begin{equation}
\min(\tdr) = (\eta \varOmega)^{-1} \approx 120 \sqrt{m_\ast} \au/f_T ~{\rm yr}\, .
\end{equation}
The ``meter-size barrier" refers roughly to the size of $\taus = 1$ bodies in the inner disk.  As \Fig{fig:ts} shows, $\sim$ cm-sized bodies drift fastest in the outer disk.  Particle inertia both lengthens the shortest drift time --- as $\min(\tdr) \propto \sqrt{\fdr} = \rho_{\rm tot}/\rho\gs$ --- and shifts it to larger $\taus = \rho_{\rm tot}/\rho\gs$.

In the small particle particle regime, with $\taus < 1$
 \begin{equation} \label{eq:tdrifttight}
{t_{\rm dr,tight}} \approx {\fdr \over 2\eta \varOmega \taus} \approx 2.5 \times 10^5 {\fdr \Fg \sqrt{m_\ast} \over a_{\rm mm} f_T \sqrt{\au}} ~ {\rm yr}\, .
 \end{equation}  
Larger particles drift faster in the test particle limit.  However inertia can cancel this trend.  With a fixed level of weak turbulence that ensures $\rho \gg \rho\gs$, \Eq{eq:rhopg} gives $\fdr \approx Z^2 \taus/\alpha$, canceling the explicit size dependence above.

For larger particles with $\taus \gtrsim 1$, the drift time increases with size and radial distance as
\begin{equation} \label{eq:tdriftloose}
t_{\rm dr,loose} \approx {\taus \over 2\eta \varOmega} \approx 2.7 \times 10^5 { a_{\rm m} \sqrt{m_\ast} \over  f_T \Fg } R_{50}^{5/2}  ~ {\rm yr} \, ,
 \end{equation}
where $R_{50} \equiv R /(50\,\rm{AU})$ and $a_{\rm m} = a/(1\,{\rm m})$.  There is no inertial correction for large solids with $\taus \gg  \rho_{\rm tot}/\rho\gs$, because drag forces are too weak to slow the headwind.

\begin{figure}[tb] 
\if\submitms y
  \includegraphics[width=4in]{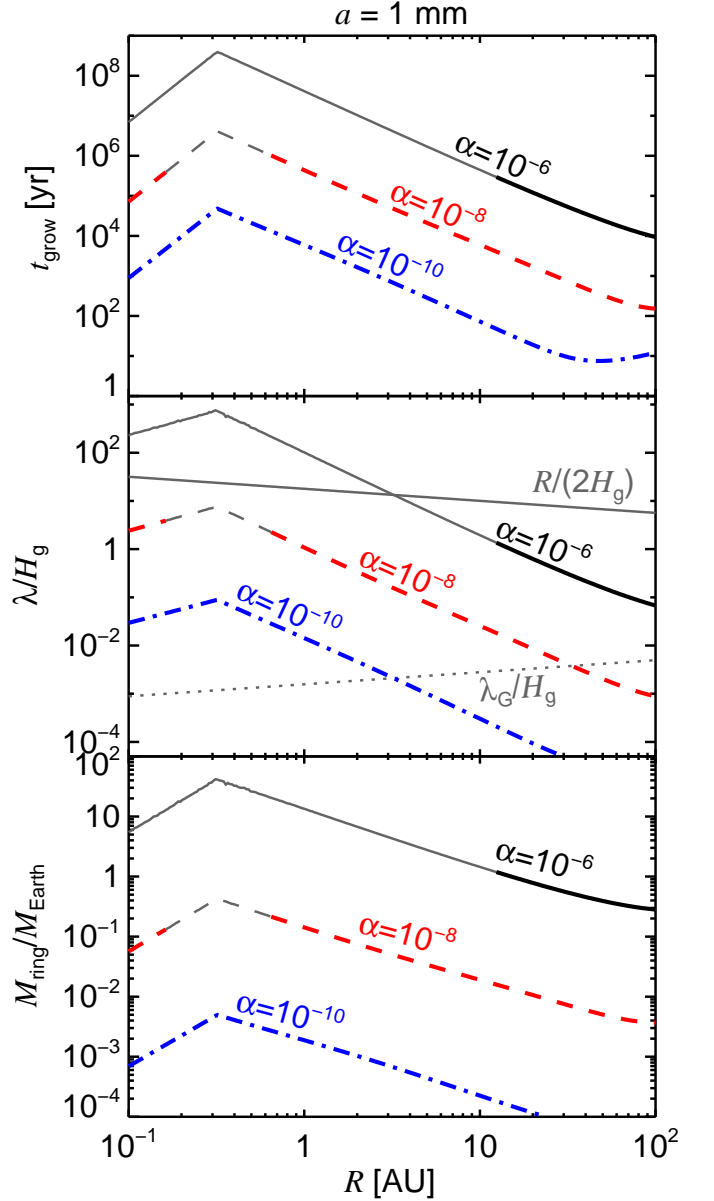} 
\else
\hspace{-1cm}
      \includegraphics[width=3.9in]{f2.eps} 
   \fi
   \caption{Properties of the dissipative gravitational collapse of mm-sized particles  vs.\ disk radius in the reference model for several values of the turbulent diffusion parameter $\alpha = 10^{-6}, 10^{-8}, 10^{-10}$ (\emph{solid, dashed, and dot-dashed curves}).  Grey sections of  curves denotes when growth is too slow.  While remarkably low, these $\alpha$ values do not preclude stronger turbulence away from the midplane, and also ignore the possibility that small-scale clumping reduces diffusion. 
   \emph{Top}: Growth times in years.
   \emph{Middle}: Wavelengths in AU compared to (half) the disk radius, the gas scale height $h\gs$, and the standard GI wavelength $\lambda_G$ (\emph{brown, black and grey dotted lines, respectively}).
   \emph{Bottom panel}: Mass of collapsing ring in Earth masses.  Subsequent fragmentation yeilds much smaller planetesimal masses. See text (\S\ref{sec:fixalpha}) for details.
   }
   \label{fig:al6810}
\end{figure}

\begin{figure}[htbp] 
\if\submitms y
   \includegraphics[width=4in]{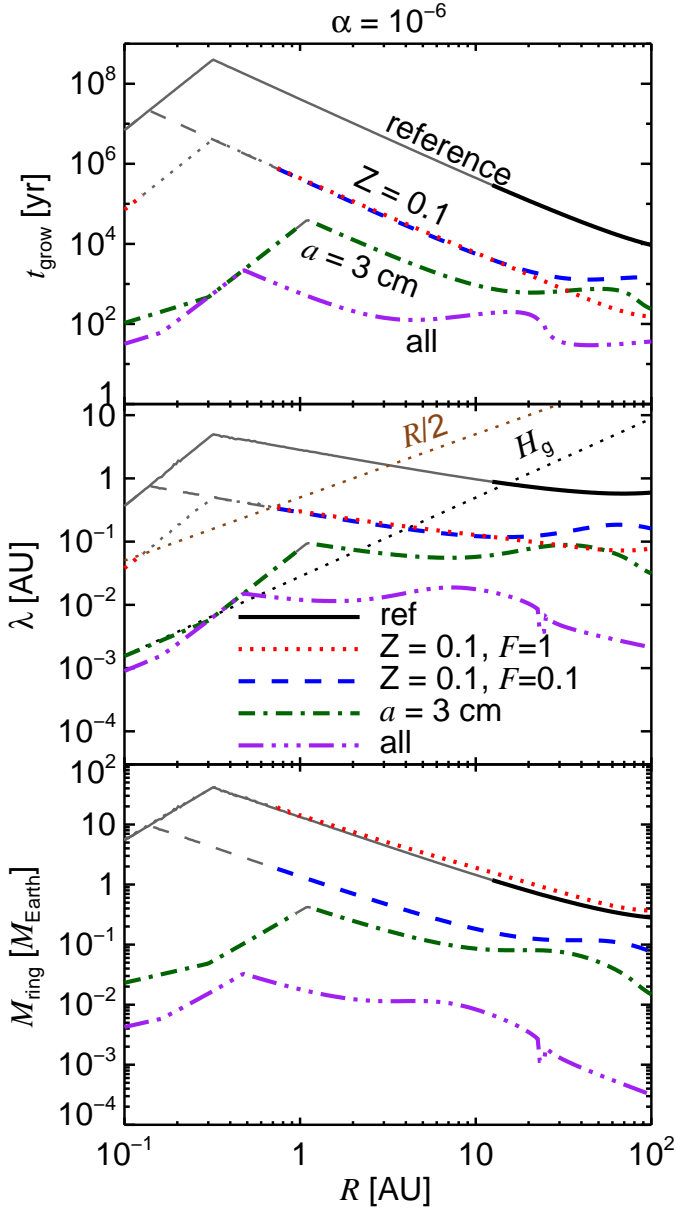} 
\else
\hspace{-1cm}
      \includegraphics[width=3.9in]{f3.eps} 
\fi
   \caption{Similar to \Fig{fig:al6810} with turbulence held fixed at $\alpha = 10^{-6}$ and other parameters varied.  The reference model with $a = 1$ mm solids (\emph{solid black curve}) is shown again for comparison.   Larger $a = 3$ cm solids (\emph{green dot-dashed curves}) give faster and smaller scale growth.  Increasing the particle-to-gas surface density ratio to $Z = 0.1$ also promotes growth, whether the gas content stays fixed ($\Fg = 1.0$, \emph{red dotted curve}) or is depleted ($\Fg = 0.1$, \emph{blue dashed curve}).  With both gas depletion and 3 cm particles (\emph{purple triple-dot-dashed curve}) growth is even faster and smaller scale.  The kink in $\lambda$ and $M_{\rm ring}$ at 20 AU is a real feature (not a numerical artifact) as explained in \S\ref{sec:QT1}.}
   \label{fig:al6var}
\end{figure}

\section{Numerical Results}\label{sec:results} 
We now present the main numerical results for the gravitational collapse of solids in a turbulent gas disk.  The fiducial protoplanetary disk models of \S\ref{sec:disk} provide the physical values.  Our most commonly plotted quantities are the growth time, wavelength and mass of solids in a collapsing ring, which we define as
\begin{subeqnarray}  
t_{\rm grow} &\equiv& [\varOmega\, \mathrm{Real}(\gamma)]^{-1}\, ,\\
\lambda &\equiv&\lambda_G/\mk \, ,\\
M_{\rm ring} &\equiv& 2 \pi \lambda R \Sigma \, .
\end{subeqnarray} 
To obtain these quantities, we solve the dispersion relation in \Eq{eq:disp} for the fastest growing modes.  This solution involves maximizing the real part of the (dimensionless) growth rate $\gamma$ with respect to the (dimensionless) wavenumber $\mk$, see \S\ref{sec:GenBehavior} for details.  We quote a ring mass because the modes are axisymmetric, but emphasize that this is \emph{not} a predicted planetesimal mass.  Gravitational fragmentation during later stages of collapse would likely produce smaller bodies. 

\subsection{When is Collapse Relevant?}\label{sec:relevant}
Since our instability always produces growth (see \S\ref{sec:proof}), we require the following conditions be met for collapse to be relevant:
\begin{subeqnarray}  \label{eq:cond}
t_{\rm grow} &<& t_{\rm drift}  \, ,\\
t_{\rm grow} &<& t_{\rm life} = 10^6~{\rm yrs} \, ,\\
\lambda &<& R/2\, .
\end{subeqnarray} 
Growth must be faster than both the radial drift time of \Eq{eq:tdrift} and an assumed $10^6$ year disk lifetime.  Also the wavelength must fit comfortably inside the disk.  The drift requirement is usually the most stringent.  

The characteristic growth time, $t_{\rm grow}$, only represents a single e-folding.  Many e-foldings are generally required for significant density growth.  The actual time to form planetesimals depends on seed perturbation amplitudes and non-linear evolution, both beyond this analysis.   To avoid introducing another uncertain parameter, we simply comment that many e-foldings are possible because growth could proceed over several $t_{\rm drift}$, especially since the growth of particle overdensities will slow the actual drift timescales.  Moreover disk lifetimes could extend to $\sim 10$ Myr, though this is less often the relevant constraint.  These issues touch on the application of a local instability to a globally evolving disk, as discussed further in \S\ref{sec:growthdisc}.

\subsection{Growth for Fixed $\alpha$}\label{sec:fixalpha}
Our reference model consists of particles with radius $a = 1$ mm in our minimum mass disk model with $\Fg = 1$ and $Z =0.01$.  \Fig{fig:al6810} plots growth times and wavelengths for different levels of the turbulent parameter $\alpha$ with all other disk quantities fixed.  Curves become light grey whenever the conditions of \Eq{eq:cond} are not met.  Low $\alpha$ values are required for collapse in our nominal disk.  With $\alpha = 10^{-6}$ collapse only proceeds outside 10 AU.  For $\alpha = 10^{-8}$ collapse occurs everywhere except a region near $R = 0.3$ AU, where the peak in growth times corresponds with the minima in stopping times  at the Epstein-Stokes transition, see \Fig{fig:ts}.  

Increasing the strength of turbulence gives slower collapse that acts over long wavelengths.   Our physical derivation of \Eqs{eq:lDapprox}{eq:sDapprox} explains the basic scalings.  Including the more precise coefficients of \Eq{eq:growI}, the growth time for $\taus \ll 1$ becomes
\begin{equation} \label{eq:tgrow}
t_{\rm grow} \simeq {1 \over \varOmega} {\alpha Q\gs^2 \over Z^2 \taus^2}\, .
\end{equation} 
The more rapid collapse in the outer disk occurs mainly due to weaker drag coupling, with larger $\taus$.  This aerodynamic effect overcomes the radial increase in orbital time scales.  Stronger self-gravity, as measured by $Q\gs$, also assists growth in the outer disk.  

Substituting numerical scalings for Epstein drag and the reference disk, the growth properties are
\begin{subeqnarray} \label{eq:DDGInum}
{t_{\rm grow} \over {\rm yr} } &\approx&  4 \times10^{5} {\alpha_{-8}f_T \sqrt{m_\ast} \over Z_\%^2 a_{\rm mm}^2 } \au^{-2}\, , \slabel{eq:tgrowI}\\
{\lambda \over {\rm AU}} &\approx& 0.028 {\alpha_{-8}  f_T \over Z_\% a_{\rm mm}  } \au^{-1/2}\, , \slabel{eq:lambdaI} \\
{M_{\rm ring} \over M_\oplus} &\approx& 0.13 {\alpha_{-8} \Fg f_T \over a_{\rm mm}\au}\slabel{eq:MringI}\, ,
\end{subeqnarray} 
where turbulence is normalized to $\alpha_{-8} = \alpha/10^{-8}$.  Different scalings (not given) apply to the Stokes regime at small $R$.

Significantly faster and smaller scale growth is possible with deviations from the reference model.   \Fig{fig:al6var} explores the effect of varying disk parameters while keeping $\alpha = 10^{-6}$ fixed.  Larger particle radii with bigger $\taus$ give much faster growth, as the  $a = 3$ cm (\emph{dot-dashed green}) curves demonstrate.  Deviations from powerlaw behavior at large $R$  arise from the transition to loose coupling, $\taus \gtrsim 1$, which  \S\ref{sec:GenBehavior} explores in more detail.  

Higher particle fractions, $Z$, produce faster growth by increasing particle self-gravity.  As an added benefit, more particle inertia slows radial drift.  Whether $Z$ increases by adding solids or depleting gas (or some combination) the effect on growth rates and  wavelengths is similar.  Indeed  \Eqs{eq:tgrowI}{eq:lambdaI} are independent of $\Fg$.   While a higher mass disk has stronger particle self-gravity (at fixed $Z$), higher gas densities cancel this effect by giving tighter coupling (in the Epstein, but not Stokes, regime).   The ring mass in \Eq{eq:MringI} does increase with $\Fg$. At fixed  $Z$, there is obviously more particle mass in a fixed area when the total mass rises.

\Fig{fig:al6var} demonstrates these metallicity effects for two high $Z = 0.1$ models, with $\Fg = 1$ (\emph{red dotted curve}) and $\Fg = 0.1$  (\emph{blue dashed curve}).  The overlap in $t_{\rm grow}$ and $\lambda$ is evident at intermediate $R$.  The curves separate at small $R$ with the transition to Stokes drag, and at large $R$ with the transition to $\taus > 1$, which occurs at smaller $R$ in the gas depleted model.

The effect of particle growth and enhanced metallicity can interact constructively.  The purple (triple dot-dashed) curve labelled ``all" shows this result by combining the $a = 3$ cm, $Z = 0.1$ and $\Fg = 0.1$ parameters.  Growth times are even faster and ring masses are much smaller when these effects are considered together.

\begin{figure}[tb!] 
\if\submitms y
 \plotone{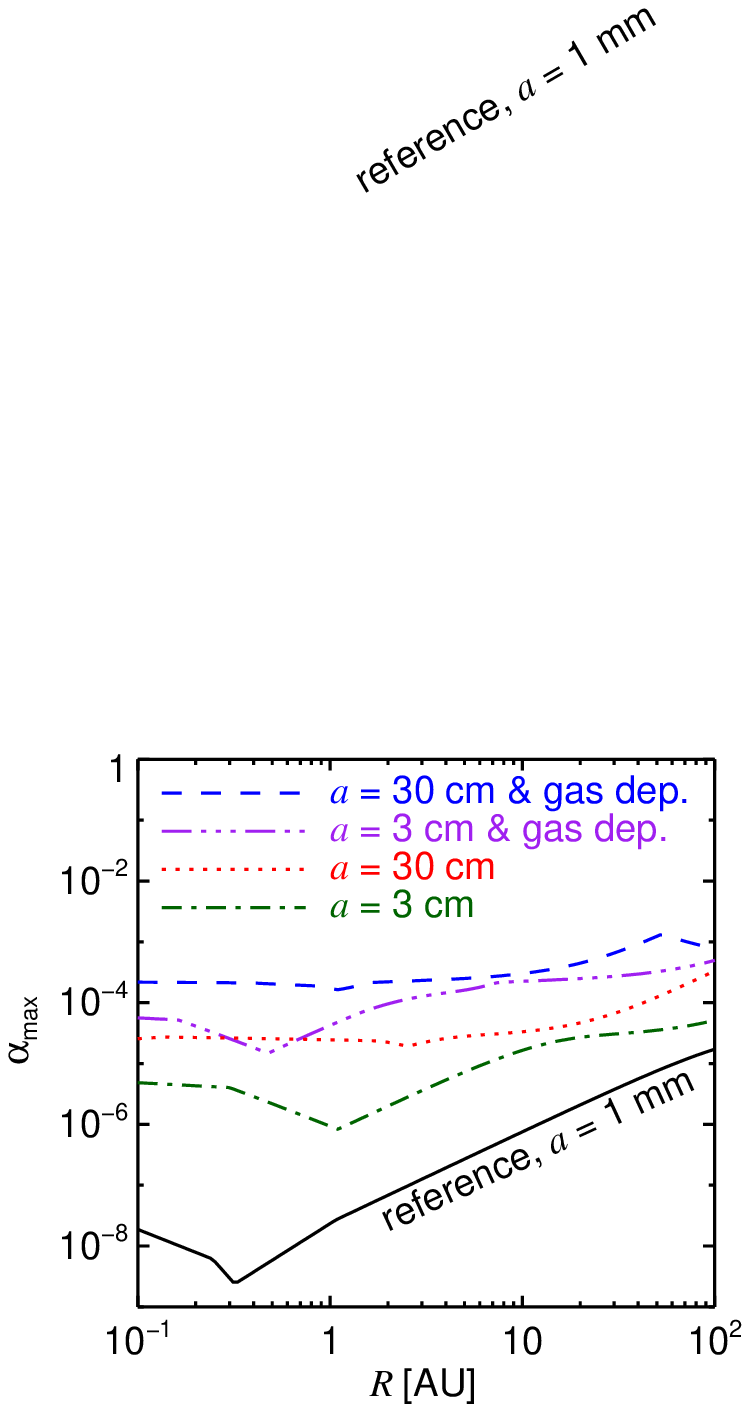}
\else
 \hspace{-1cm}
      \includegraphics[width=3.9in]{f4.eps} 
\fi   
   \caption{The maximum strength of turbulence, given by $\alpha_{\rm max}$, which allows collapse.   The criteria for sufficiently rapid collapse are described in \Eq{eq:cond}.  The $\alpha_{\rm max}$ values increase with particle size as shown for $a = 1$ mm (\emph{black solid curve}), $a = 3$ cm (\emph{dash-dotted green curve}) and $a = 30$ cm (\emph{dotted red curve}).  The enrichment of the solid to gas ratio also increases $\alpha_{\rm max}$.  We show this for gas depleted models with $Z = 0.1$ and $\Fg = 0.1$ for both $a = 3$ cm (\emph{triple-dot dashed purple curve}) and $a = 30$ cm (\emph{dashed blue curve}).  See \Fig{fig:almaxtlm} for the properties of modes with $\alpha = \alpha_{max}$.}
   \label{fig:almax}
\end{figure}

\begin{figure}[tb!] 
\if\submitms y
 \includegraphics[width=4in]{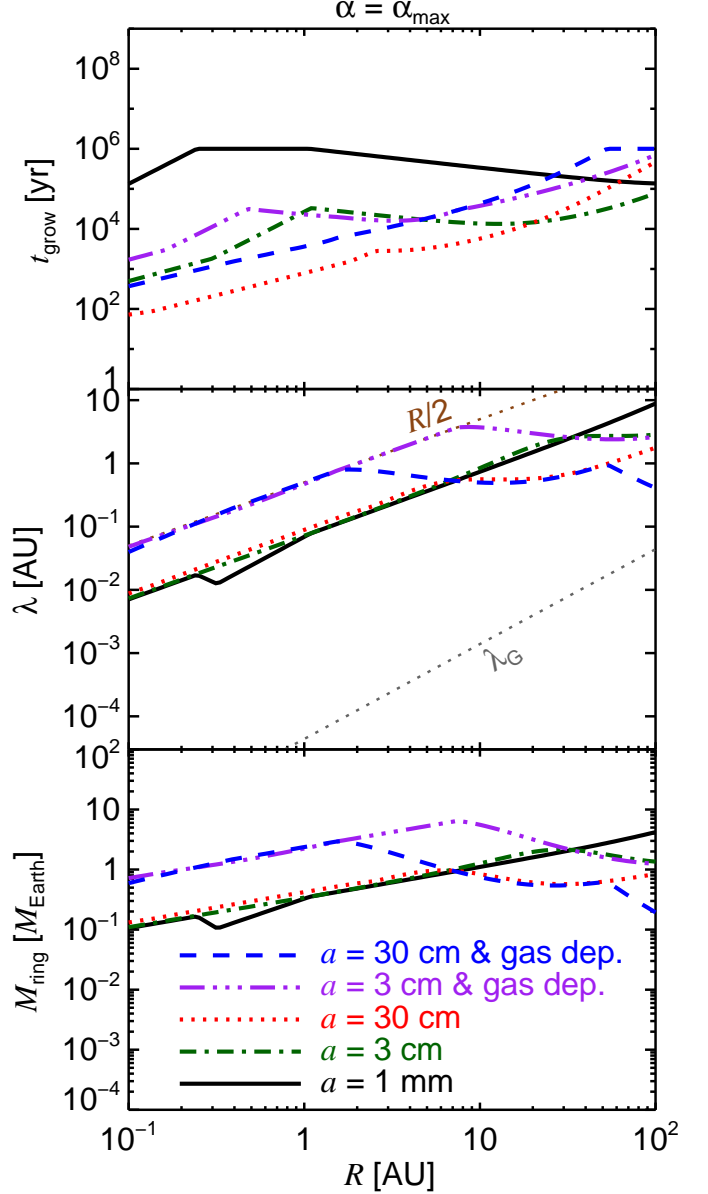}
  \else
\hspace{-1cm}
      \includegraphics[width=3.9in]{f5.eps} 
\fi
   \caption{Properties of modes with the maximum allowed turbulence, $\alpha = \alpha_{\rm max}$, for the same models as in \Fig{fig:almax}. We can visually identify which constraint sets $\alpha_{\rm max}$: if $t_{\rm grow} = 10^6$ yr, it is the disk lifetime, and when  $\lambda = R/2$, it is the disk size.  Otherwise the drift time scale is the relevant constraint.  See the text (\S\ref{sec:alphamax}) for an explanation of the overlapping wavelengths and masses.}
   \label{fig:almaxtlm}
\end{figure}

\begin{figure}[tb!] 
\if\submitms y
 \includegraphics[width=6in]{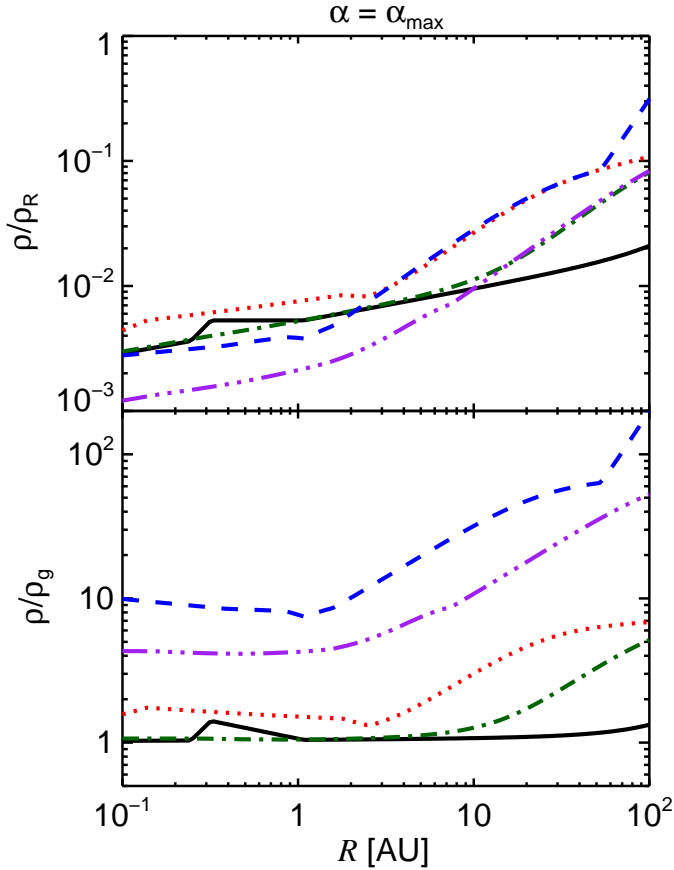}
  \else
\hspace{-1cm}
      \includegraphics[width=3.9in]{f6.eps} 
\fi
   \caption{The minimum particle density, $\rho$, that gives collapse, obtained by setting $\alpha = \alpha_{\rm max}$, using the same disk models as in \Figs{fig:almax}{fig:almaxtlm}.  (\emph{Top}): $\rho$ plotted relative to the Roche density [\Eq{eq:rhoR}], demonstrating that secular GI occurs well below this standard threshold. (\emph{Bottom}): $\rho$ plotted relative to gas density to show that drag feedback is always significant.
}
   \label{fig:almaxrho}
\end{figure}

\subsection{Maximum Allowed Turbulence}\label{sec:alphamax}
Instead of fixing the level of turbulence, we now determine the strongest level of turbulence that allows collapse.  \Fig{fig:almax} plots $\alpha_{\rm max}$, the largest $\alpha$ value for which growing modes satisfy all the constraints of \Eq{eq:cond}.   Not surprisingly, the same factors that give faster growth --- larger solids and higher particle mass fractions --- also yield larger $\alpha_{\rm max}$.  

The constraints on turbulence are strongest inside a few AU, where $\alpha \lesssim 10^{-4}$ is required for sizes up to 30 cm, even with some metallicity enrichment.  While the outer disk generally allows collapse with stronger turbulence, it still requires $\alpha \lesssim 10^{-3}$.  

Analytic fits to $\alpha_{\rm max}$ come with even more caveats than usual, but are useful for a rough understanding.  When drift speeds are the limiting constraint, the condition $t_{\rm grow} < \tdr$ sets
\begin{equation} \label{eq:alphadrift}
\alpha_{\rm max} \simeq {\taus Z^2 \fdr \over 2 \eta Q\gs^2} \approx 0.6 \times10^{-8} {a_{\rm mm} Z_\%^2 \Fg \fdr \over f_T^2} \au^{3/2}\, ,
\end{equation} 
for $\taus \ll 1$.  (Again the numerical scalings apply only to Epstein drag.)  This basic scaling explains many of the curves in  \Fig{fig:almax}, e.g.\  the $a = 1$ mm curve beyond 1 AU and segments of the $a = 3$ cm curves.  For transitional $\taus \sim 1$ values,  $\alpha_{\rm max}$ has a flat $R$ dependence, e.g.\ for the $a = 3$ cm curve beyond 10 AU.  For $\taus \gg 1$, a similar constraint to \Eq{eq:alphadrift} holds, again with $\alpha_{\rm max} \propto R^{3/2}$ as seen in the $a = 30$ cm curve beyond 20 AU.  The kink near 60 AU in the gas depleted $a = 30$ cm curve occurs because the disk lifetime constraint becomes relevant.


The upper limit to allowed turbulence, $\alpha_{\rm max}$, varies greatly with disk model and has no absolute upper limit. The  range of models we consider all require $\alpha_{\rm max} \lesssim 10^{-3}$, and lower values in the inner disk.

\subsubsection{Mode Properties}
To better understand these limits on turbulence, we plot the properties of modes with $\alpha = \alpha_{\rm max}$ in \Fig{fig:almaxtlm}.  These growth times, wavelengths and masses here take  the largest allowed values for a given disk model.  For $\alpha < \alpha_{\rm max}$, collapse is more rapid and smaller scale. 

\Fig{fig:almaxtlm} allows us to readily identify the relevant constraint that sets $\alpha_{\rm max}$.  The $10^6$ year lifetime constraint applies when the drift time scale is even longer.  As shown in the top panel, this occurs both for loose coupling (beyond 60 AU in the gas depleted $a = 30$ cm model)  and for very tight coupling (mm-sized solids near $\sim 0.3$ AU at the Stokes-Epstein transition).  In the middle panel, we see that the disk size constrains wavelengths in our enhanced metallicity models, especially in the inner disk. 
The radial drift time scale is the relevant constraint on $\alpha_{\rm max}$ wherever $t_{\rm grow} \neq 10^6$ yr and $\lambda \neq R/2$.

A striking feature of \Fig{fig:almaxtlm} is the clustering of the wavelength (and also mass) curves for modes for models with very different $\taus$.  Substitution of  $\alpha_{\rm max} \propto \taus$ from \Eq{eq:alphadrift} into the wavelength result of \Eq{eq:growI} gives
\begin{equation} \label{eq:lambdamax}
\lambda(\alpha_{\rm max}) \approx {\fdr \over 2 \eta} \lambda_G \approx 0.017 {\fdr \Fg Z_\% \au \over f_T}~{\rm AU}\, ,
\end{equation}
which is independent of $\taus$.  Thus the maximum wavelength imposed by drift is at least $1/(2 \eta)$ larger than the standard GI wavelength, and more if the inertial correction factor, $\fdr$, is significant.  The gas depleted curves in \Fig{fig:almaxtlm} have larger maximum wavelengths when $\taus < 1$.  In these cases, higher particle inertia increases $\fdr$ values, raising $\lambda(\alpha_{\rm max})$ to the $\lambda = R/2$ limit.

\subsubsection{Minimum Particle Density}\label{sec:rhop}
The volume density of solids, $\rho$, is both a measure of self-gravity --- though an indirect and imprecise one for secular GI ---  and a measure of feedback effects that give rise to drag instabilities.  When $\alpha = \alpha_{\rm max}$ we find the minimum $\rho$ required for collapse.  

 The top panel of \Fig{fig:almaxrho} plots $\rho/\rho_{\rm R} = 0.2/\QR$.  Since $\rho \ll \rho_{\rm R}$, secular GI occurs well below this standard threshold.  While \S\ref{sec:proof} proved that collapse formally occurs for any $\QR$ and thus $\rho$, these numerical results demonstrate collapse for small $\rho$ when astrophysical constraints are imposed.    Below the $\rho$ values plotted in \Fig{fig:almaxrho}, collapse is too slow to be relevant.
 
We emphasize that density is an indirect measure of the relevance of secular GI.  Radial diffusion is the dominant stabilizing influence, so the vertical diffusion that sets $\rho$ is more of a side effect.  If we consider asymmetric turbulence and vary just the vertical diffusion,  $\rho$ would change, but growth rates would be unaffected because the finite thickness correction is negligible for long wavelength modes.  Drift rates however would change from the inertial correction. 

The bottom panel of \Fig{fig:almaxrho} plots the particle density --- again, the minimum value required for collapse --- relative to the gas density.  Remarkably, the minimum mass models with $\Fg = 1.0$ all have $\rho \gtrsim  \rho\gs$.   Applying the drift constraint of \Eq{eq:alphadrift} to the density relation in \Eq{eq:rhopg} gives
\begin{equation} \label{eq:feedback}
{\rho (\alpha_{\rm max})\over \rho\gs} \approx Q\gs \sqrt{2\eta \over \fdr} \approx 2.0 {f_T \over \Fg \sqrt{\fdr}}\, .
\end{equation} 
Thus $\rho/\rho\gs$ is unity for $\Fg = 1$ models (when $\taus \lesssim 1$) and larger for the gas depleted $\Fg = 0.1$ models, explaining the bottom panel \Fig{fig:almaxrho}.  It is a numerical coincidence ---  unrelated to the mechanism of the instability --- that standard minimum mass gas disks give $Q\gs \sqrt{2 \eta}$ of order unity.  But this coincidence is significant.  The consequence that $\rho \gtrsim \rho\gs$ for gravitational collapse means that drag feedback is relevant, and can drive particle-driven turbulence.  See \S\ref{sec:growthdisc} for more discussion.

\begin{figure*}[tb!] 
\if\submitms y
\vspace{1in}
   \plotone{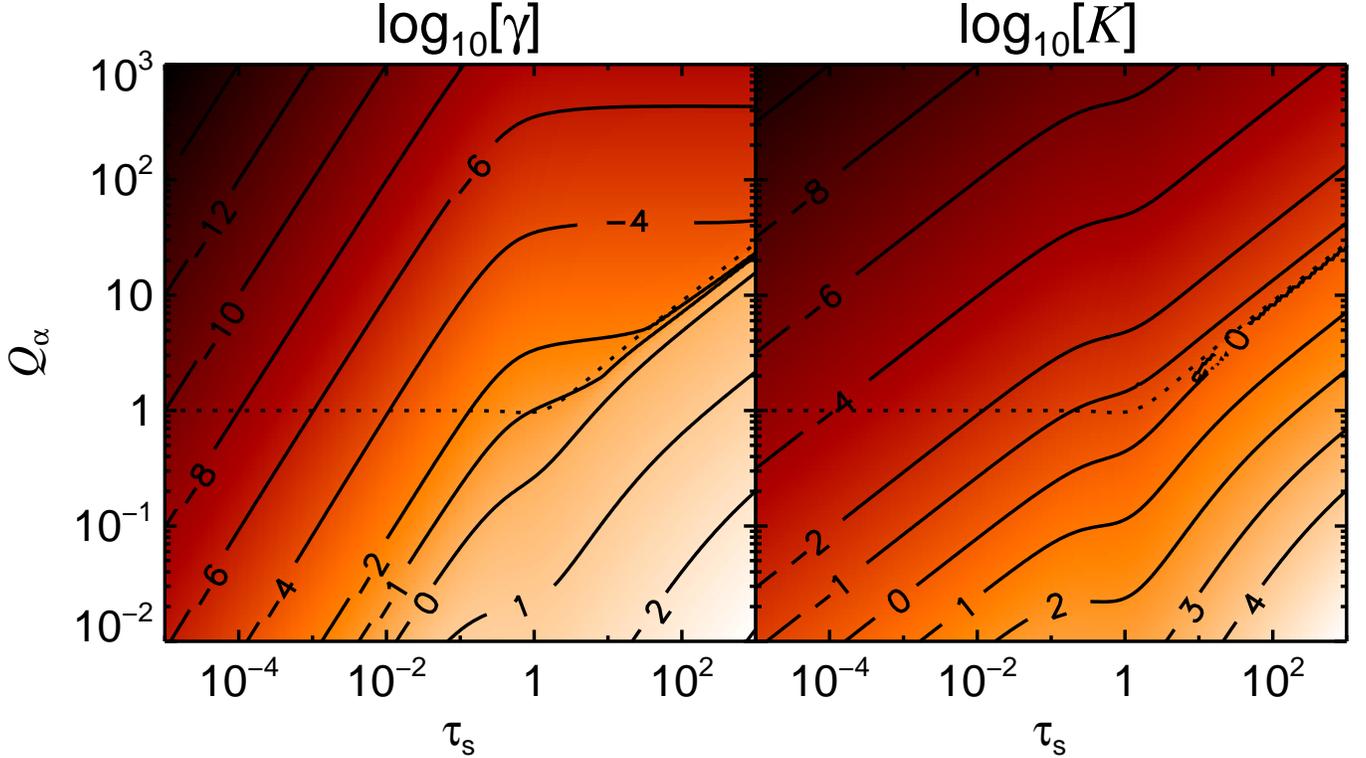}
   \else
\hspace{-1cm}
      \includegraphics[width=7.75in]{f7.eps} 
 \vspace{-.75cm}
\fi
   \caption{(\emph{Solid curves and grayscale}): Contours of (the log of) the dimensionless growth rate, $\gamma$, (\emph{left panel}) and wavenumber, $\mk$, (\emph{right panel}) of the fastest growing mode as a function of  the strength of the turbulence -- measured by $\QDg$ --  and the size of particles -- measured by $\taus$.  The \emph{dotted curve} denotes $\QT = 1$, the nominal threshold for standard GI.   Slow growth occurs for $\QT > 1$ (above the dotted curve) and the $\QT = 1$ contour only has significance for $\taus \gg 1$ .  See  \Tab{tab:region} and \S\ref{sec:GenBehavior} for explanations of different parameter regimes.}
   \label{fig:QaTs}
\end{figure*}

\begin{figure}[b!] 
\if\submitms y
\plotone{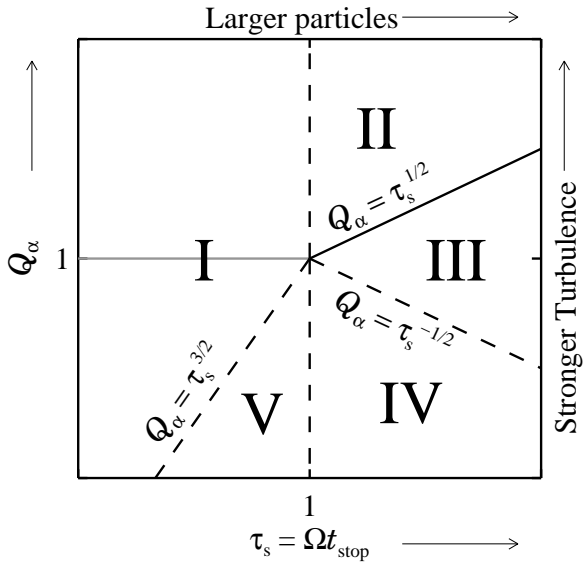}
\else
 \hspace{-1cm}
      \includegraphics[width=3.5in]{f8.eps} 
\fi   
   \caption{Overview of $\QDg$ vs.\ $\taus$ parameter space.  Region labels I --- V are used in \Tab{tab:region} and the text.  Regions can be identified in  \Fig{fig:QaTs}, which is not centered on (1,1).  The solid line that divides region I and runs between regions II and III) denotes $\QT = 1$, marginal stability for standard GI.  }
   \label{fig:QaTsSpace}
\end{figure}

\section{Growth Across Parameter Space}\label{sec:GenBehavior}
We now survey the behavior of our instabilities across  the full range of parameter space.  Varying only two parameters --- the stopping time, $\taus$, and the strength of turbulence relative to particle self-gravity, $\QDg$ --- covers all possibilities, thanks to the turbulence model of \S\ref{sec:turbmodel}.  Once solutions to the dispersion relation of \Eq{eq:disp} are tabulated,  translating the conditions in an arbitrary disk model to appropriate values of $\taus$ and $\QDg$ allows simple look-up of the solutions.
Furthermore, we gain both a deeper understanding and a useful check on numerical solutions by understanding the analytic behavior for extreme values of the input parameters.

\Fig{fig:QaTs} maps the growth rate, $\gamma\fgm$, and wavenumber, $\mk\fgm$, of the fastest growing modes versus $\QDg$ and $\taus$.   We obtain these solutions by maximizing each of the three roots of \Eq{eq:disp} with respect to $\mk$ over a well-sampled grid of  $\taus$ and $\QDg$ values.\footnote{For extreme values of the input parameters, renormalizing $\mk$ towards a known solution as $\mk \rightarrow \mk' = \mk \QDg^2/ \taus$ helps avoid numerical difficulties.}   

\Fig{fig:QaTs} shows the general trend that both growth rates and wavenumbers tend to increase for small $\QDg$ (weaker turbulence or stronger self-gravity) and larger $\taus$ (bigger solids or weaker drag).  However the behavior has a different character in different regions of parameter space.  We identify five distinct regions of parameter space, shown in \Fig{fig:QaTsSpace}.   \Tab{tab:region} lists the analytic growth rates and wavenumbers by region. 

We now describe the behavior by regions, with less relevant cases deferred to appendix \ref{sec:regapp}.  The transition from secular to standard GI occurs across the region II to III boundary.  \Fig{fig:QaTs} shows the corresponding sharp rise in growth rates, explored further in \S\ref{sec:QT1}.

\begin{deluxetable}{@{} ccccc @{}}
  
   \tablecaption{Growth across parameter space} 
  \tablehead{
  \# & Drag & Turbulence & $\gamma $ & $\mk$ }
      \startdata
           I      &$\taus \ll 1$ & $\QDg \gg \taus^{3/2}$ &$\taus^2 /\QDg^2$ & $\taus /\QDg^2$ \\
            II      &$\taus \gg 1$ & $\QDg \gg \sqrt{\taus}$ & $0.2 /\QDg^2$ & $0.2 \taus /\QDg^2$ \\
               III      &$\taus \gg 1$ & ${1 \over \sqrt{\taus}}  \ll \QDg \ll \sqrt{\taus}$ & $0.7 \sqrt{\taus}/\QDg$ & $0.5 \taus / \QDg^2 $\\
  IV      &$\taus \gg 1$ & $\QDg \ll 1/\sqrt{\taus}$ & $0.6 \left({\taus \over \QDg}\right)^{2 \over 3}$ & $0.4\left({\taus \over \QDg}\right)^{4 \over 3}$ \\
  V      &$\taus \ll  1$ & $\QDg \ll \taus^{3/2}$ & $ \QDg^{-2/3}$ & $\QDg^{-4/3}$ \\  
          \enddata
  \tablecomments{Col.\ (1): Region label, see also \Fig{fig:QaTsSpace}. Cols.\ (2) \& (3): Region boundaries in $\taus$-$\QDg$ space. Cols.\ (4) \& (5):  Growth rate and wavenumber of fastest growing modes.}
   \label{tab:region}
\end{deluxetable}

\subsection{Region I: Secular, Tight}
Region I  is the most relevant for planetesimal formation.  With $\taus \ll 1$ and $\QDg \gg \taus^{3/2}$, it describes tight coupling and turbulence that is not incredibly weak.\footnote{Appendix \ref{sec:regV} shows that the weak turbulence solutions in region V are not relevant.}   The physical derivation of secular GI  in \S\ref{sec:OOM} --- a simple balance between terminal velocity infall and diffusion ---applies in region I.
  
The fastest growth rate and corresponding wavenumber in region I,
\begin{equation}\label{eq:growI}
\gamma_{\rm I} = \taus^2/\QDg^2 \, ,\mk_{\rm I} =  \taus/\QDg^2 \, , 
\end{equation} 
are limiting solutions of the full dispersion relation.  They are more exact versions of  the physically derived \Eqs{eq:lDapprox}{eq:sDapprox}.   The smooth variation with $\QDg$, also evident in \Fig{fig:QaTs}, demonstrates the insignificance of the $\QDg = \QT = 1$ transition for $\taus \ll 1$.  

Growth rates in region I can be as fast as $\gamma \sim 1/\taus \gg 1$ at the region V transition.  Nevertheless, growth rates are slower than dynamical, $\gamma \ll 1$,  because our disk models have $\QDg > 1$ as in \Eq{eq:QDgnum}.

The thin disk approximation holds in region I.  From \Eqs{eq:QRtightapp}{eq:growI}, $\mk_{\rm I} \QR \ll1$ for weak self-gravity with $\QDg > \sqrt{\taus}$.  The strong self-gravity case of $\QDg < \sqrt{\taus} \ll 1$, is less relevant, but still gives only a modest finite thickness correction, $\mk_{\rm I} \QR \sim1$.  (And thankfully we need not sub-divide region I.)

 Density waves are rapidly damped by drag in region I.  Growth occurs via the neutral mode, introduced in \S\ref{sec:standardGI} as trivial in the absence of drag.  The eigenfunctions of the particle response are $\tilde{U}_{\rm I} =  2 \tilde{V}_{\rm I}/\taus = 2\imath \taus \tilde{\sigma}_{\rm I}$.   This strong density response, with $\tilde{\sigma} \gg \tilde{U} \gg  \tilde{V}$, makes it unlikely that secondary flow instabilities will halt the gravitational collapse once it begins.

Wavenumbers other than the fastest growing can also collapse, at a rate determined by the leading order dispersion relation:
\begin{equation}\label{eq:dispI} 
\gamma_{\rm I}(\mk)  \approx 2\taus \mk - \mk^2 \QDg^2 \, . 
\end{equation} 
This result has implications for how collapse proceeds for a distribution of particle sizes as discussed in \S\ref{sec:valid}.

\subsection{Region II: Secular, Loose}\label{sec:regII}
Region II describes larger, loosely coupled solids with $\taus \gg 1$ and relatively weak self-gravity characterized by Toomre's $\QT = \QDg/\sqrt{\taus} \gg 1$.  Like the tight coupling case of region I, gravitational collapse involves the secular destabilization of a neutral mode. 

The physical derivation of region II  growth follows the qualitative description of secular GI in the introduction of \citet{gp00}.   As in \S\ref{sec:OOM} we consider a ringlike perturbation of the solids with radial self-gravity $g_r \sim G \sigma$.  The orbital speed at the outer (inner) edge of the ring increases (decreases, respectively) by $v \sim g_r/(2 \varOmega)$ in response.  Assuming the gas is unaffected by this acceleration (see \S\ref{sec:prev}), the extra headwind (tailwind for the inner edge) causes a radial drift towards the ring center.  By angular momentum conservation this drift speed is
\begin{equation} 
u \sim {2 v \over \varOmega \ts} \sim {G \sigma \over \varOmega^2 \ts}\, .
\end{equation} 
Due to the weaker coupling this scaling differs from \Eq{eq:driftapp}.  The collapse rate again follows from mass continuity as
\begin{equation} 
s \sim {G \varSigma \over \varOmega^2 \ts \lambda}\, .
\end{equation} 

As in \S\ref{sec:OOM}, we consider the minimum $\lambda$ allowed by pressure and diffusive stabilization.  It turns out that both are of roughly equal importance for loose coupling (and our turbulent stirring model).  Pressure stabilization sets a minimum wavelength
\begin{equation} \label{eq:lpII}
\lambda_P \sim {c^2 \over G \varSigma} \sim  {\ve^2 \over G \varSigma \taus} \, ,
\end{equation} 
where we use the result $c \sim \ve/\sqrt{\taus}$ from \Eq{eq:c} for the random speed of $\taus \gg 1$ solids stirred by turbulence.  Mass diffusion imposes a time scale constraint $s \lambda^2/D\ps > 1$ that also sets a minimum wavelength
\begin{equation} 
\lambda_D \sim {\varOmega D\ps \taus \over G \varSigma} \sim {\varOmega D\gs \over G \varSigma \taus} \sim {\ve^2 \over G \varSigma \taus} \, ,
\end{equation} 
using  $D \sim D\gs/\taus^2$ for the turbulent diffusion of large bodies, see \Eq{eq:D}.  The agreement with \Eq{eq:lpII} shows that pressure and diffusive stabilization are of roughly equal importance.

The resulting growth rate,$s_{\rm II} \sim \varOmega/\QDg^2$, is independent of stopping time.  As $\taus$ increases, shorter wavelengths collapse, but the rate is unchanged because there is dissipation to drive infall.  In dimensionless units the fastest growing modes in region II obey
\begin{equation}\label{eq:secloose}
\gamma_{\rm II} = \mk_{\rm II}/\taus = c_k/\QDg^2 \, ,\\
\end{equation} 
where the numerical coefficient $c_k = 4/21$ reflects the combined effects of pressure and diffusive stabilization from an expansion of the full dispersion relation. Aside from this numerical coefficient, the wavenumber has the same basic scaling $\mk \sim \taus/\QDg^2$ in regions I and II.

In region II, growth rates are always slower than dynamical with $\gamma_{\rm II} \ll 1$.  Furthermore, $\gamma_{\rm II} \taus \ll 1$, so that drag can dissipate excess energy generated during collapse. The thin disk approximation holds with $\mk_{\rm II} \QR \sim {\sqrt{\taus}/\QDg} \sim 1/\QT \ll 1$.  

\begin{figure*}[tb] 
\if\submitms y
   \plotone{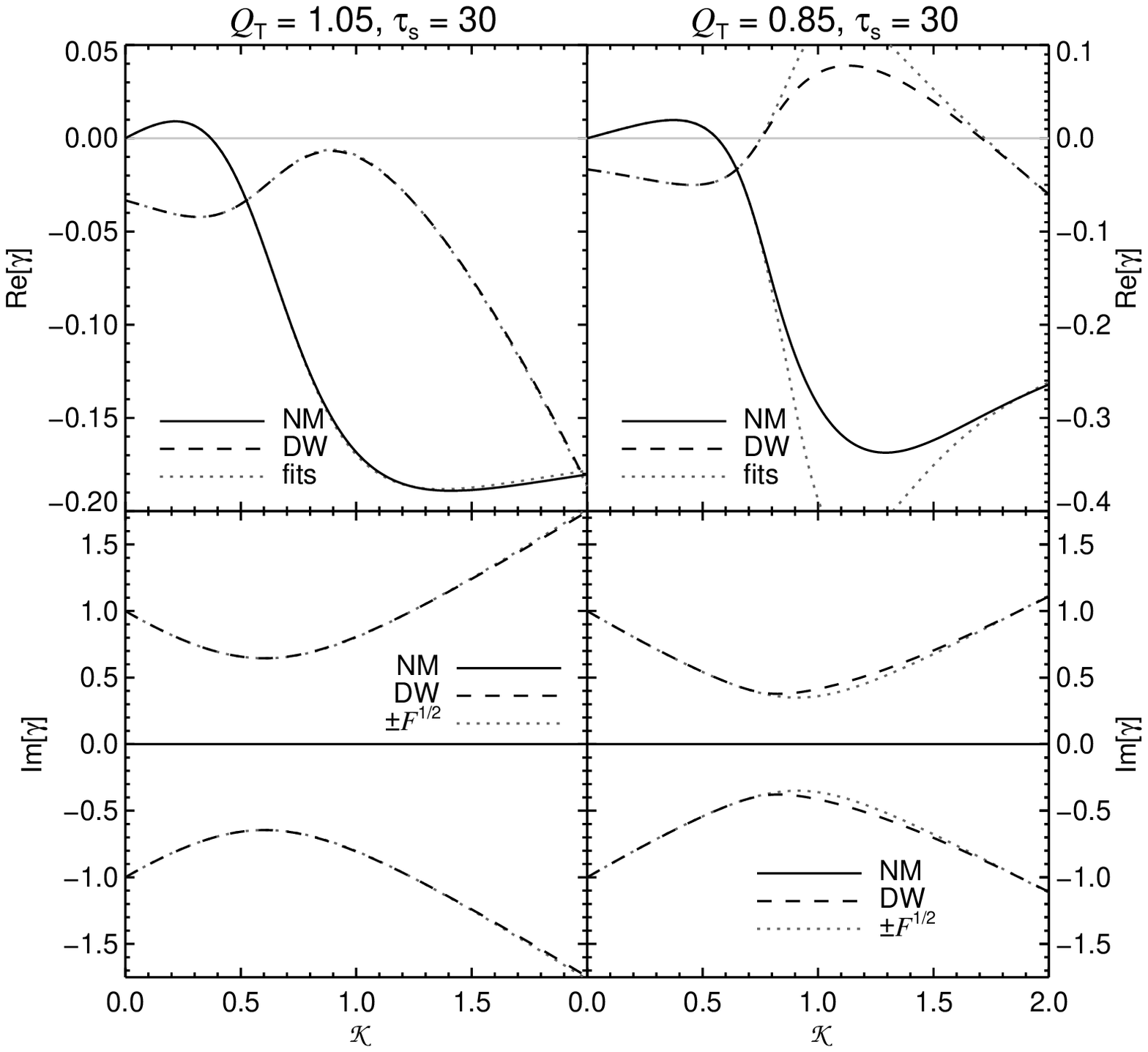}
\else
   \centering
         \includegraphics[width=7.5in]{f9.eps} 
\fi
   \caption{Growth rates \emph{(top panels)} and wave frequencies \emph{(bottom panels)}, (in orbital units)  vs.\ wavenumber $\mk$ for weak drag ($\taus = 30$) and Toomre parameters: $\QT = 1.05$ \emph{(left panels)} and $\QT = 0.85$ \emph{(right panels)}.  The zero frequency neutral mode (NM, \emph{solid curves}) is unstable at small $\mk$.   Density waves (DW, \emph{dashed curve}) are damped for $\QT = 1.05$, but exhibit growth when $\QT = 0.85$ for $\mk$ near unity.  Since $\mF>0$ (\emph{grey dotted curves, bottom panels})  all DW solutions would be stable in the absence of gas drag ($\QT = 0.85$ is stable for standard GI due to a finite thickness correction).  Analytic fits to the growth rates (\emph{dotted curves, top panel}) from a $\taus \gg 1$ expansion work well except near $\mF \approx 0$.}
   \label{fig:nm_vs_dw}
\end{figure*}

\begin{figure}[tb] 
\if\submitms y
  \includegraphics[width=4in]{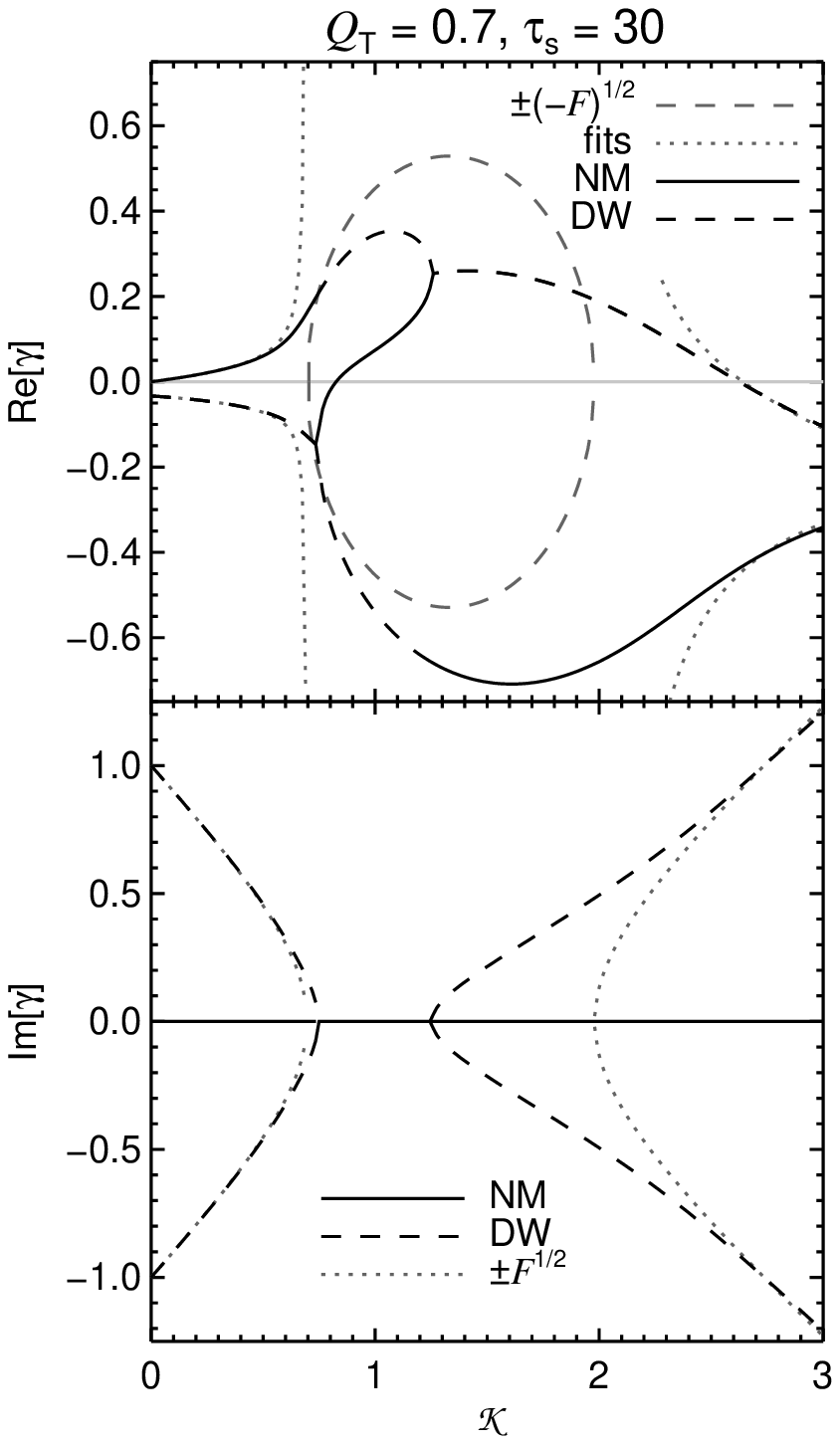} 
\else
   \vspace{.5cm}
      \includegraphics[width=3.5in]{f10.eps} 
\fi
   \caption{Similar to \Fig{fig:nm_vs_dw}, but with $\QT = 0.7$.  Stronger self-gravity gives $\mF < 0$ between $0.7 \la \mk \la 2.0$.   Without drag, standard GI gives a growth rate $\pm \mathrm{Real}(\sqrt{-\mF})$ in this region (\emph{grey dashed curves, top panel}).  Gas drag affects the growth rates and introduce two bifurcations.  At the first bifurcation (near $\mk \approx 0.7$), the unstable NM becomes an unstable DW, while the damped DW pair splits into a NM and a DW.  At the second bifurcation (where $\mk \approx 1.2$, the NM and unstable DW merge into an overstable DW pair, while the damped DW transitions to a damped NM.} 
   \label{fig:loosedyn}
\end{figure}

\subsection{Region III: Recovery of Standard GI}\label{sec:regIII}
Region III is where our model reproduces standard GI due to unstable density waves as in \S\ref{sec:standardGI}.  Radial mass diffusion is negligible  and drag forces (still relevant in region II despite $\taus \gg 1$) are weak compared to particle inertia.  Turbulence still sets velocity dispersions (giving $\QT$) and the particle layer thickness (giving $\QR$).  The boundaries of region III are $1/\sqrt{\taus} \ll \QDg \ll \sqrt{\taus}$, which requires $\taus \gg 1$.

Since $\QT \approx   \QDg /\sqrt{\taus} \ll 1$, region III is likely not accessible in practice.  Self-gravitating disks likely self-regulate to $\QT \approx 1$ \citep{glb65II,shuv2,gam01}.  Thus the boundary between regions II and III (described below) is likely more relevant than the heart of region III.   Unlike regions I and II, GI might not result in collapse in region III.  \citet{gam01} shows that energy dissipation must occur on orbital time scales for standard GI to give collapse.
With $\taus \gg 1$, dissipation by gas drag is not sufficiently rapid.

Linear growth in region III occurs at the free-fall rate
\begin{equation} \label{eq:ff}
s_{\rm ff} \sim \sqrt{G \varSigma/\lambda}\, .
\end{equation} 
The minimum wavelength is set by effective pressure support, $\lambda \lesssim \lambda_P$, from \Eq{eq:lpII}, which gives a growth rate $s_{\rm III} \sim \varOmega/\QT$.  Diffusive support is too small scale.  Long wavelengths are centrifugally stabilized if growth is slow compared to an epicyclic oscillation.  Collapse requires $s_{\rm ff} \gtrsim \varOmega$, \ie $\lambda \lesssim G \varSigma/\varOmega^2$.  The joint wavelength constraint $c^2/(G\varSigma) \lesssim \lambda \lesssim G \varSigma/\varOmega^2$ gives the Toomre instability criterion $\QT \lesssim 1$, that is satisfied throughout region III.

The fastest growing modes in region III obey
\begin{equation} 
\gamma_{\rm III} \simeq 0.7 \sqrt{\taus}/\QDg \, , \, \mk_{\rm III} = 0.5 \taus/\QDg^2\, ,
\end{equation} 
including finite thickness effects.  Since $\gamma_{\rm III} \gg 1$, collapse is faster than an orbital time scale.

\subsection{Across the $\QT \approx 1$ Transition}\label{sec:QT1}
The boundary between regions II and III has $\QT \approx 1$, which regains significance for $\taus \gg 1$.   The sharp increase in growth rates across the region  II to III boundary (seen in \Fig{fig:QaTs})  is hardly surprising since standard GI stabilizes here.  The presence of gas drag guarantees instability and gives complex behavior.   Practical interest in this region is restricted to late stage, possibly second generation, planetesimal formation in  gas depleted disks. 

 \Fig{fig:nm_vs_dw} demonstrates that both the neutral mode (NM) and a density wave (DW) can collapse, competing for the fastest growth.  For $\QT = 1.05$ only the NM grows and both DW solutions are damped at all wavelengths.  For stronger self-gravity, DW growth appears at shorter wavelengths.  By $\QT = 0.85$, the DW branch has faster growth.   This competition generates wavelength discontinuities when tracking the fastest growth.  In \Fig{fig:al6var}, the 3 cm gas depleted (\emph{purple triple-dot dashed}) curve shows such a switch near 20 AU.  In \Fig{fig:QaTs} (\emph{left panel}) this switching is evident in the winding $\mk = 1$ curve (labelled ``0" in log units).

\Fig{fig:loosedyn} demonstrates even more complex behavior when the self-gravity function $\mF$ turns negative, \ie unstable to standard GI.  Bifurcations appear where NM and DW modes suddenly change character.  These intriguing features reveal interesting dynamical behavior and suggest that care is required in numerical searches for fastest growing modes.

\subsection{Marginal Coupling}\label{sec:marg}
We now consider the special case of $\taus = 1$, when orbital and drag times are identical.   This case is of special interest because drift times are fastest and many particle concentration mechanisms are most efficient (CY10, \citealp{houches10}).  Furthermore numerical simulations are easier (not easy) without separate time scales.  
When $\taus = 1$ and $\QDg \gg 1$ (\ie the boundary between regions I and II) the limiting dispersion relation for the NM is
\begin{equation} 
\gamma_{\rm marg} \approx \mk - \left(\QDp^2 + {\QT^2 \over 2} \right)\mk^2 \approx \mk - {\QDg^2 \over 4 c_m} \mk^2
\end{equation} 
indicating stabilization by both effective pressure and diffusion as in Region II.  The numerical coefficient, $c_m = 8/65$, follows from our turbulence model [\Eq{eq:Qs}], which should not be trusted to high precision due to complex dynamics in the $\taus \approx 1$ regime.  The fastest growing modes obey $\gamma_{\rm marg} = \mk_{\rm marg}/2 = c_m/ \QDg^2$.

Physical values for the growth of $\taus = 1$ modes are
\begin{subeqnarray} 
{t_{\rm grow, marg} \over {\rm yr} }& \approx&  2.1 \times10^{3} {\alpha_{-4}f_T \sqrt{m_\ast} \over Z_\%^2 \Fg^2 } \au\, , \\
{\lambda_{\rm marg} \over {\rm AU}} &\approx& 0.29 {\alpha_{-4} f_T \over Z_\% \Fg  } \au \,\\ 
{M_{\rm ring, marg} \over M_\oplus} &\approx& 1.36 \alpha_{-4} f_T \sqrt{\au} \, ,
\end{subeqnarray} 
with $\alpha_{-4} = \alpha/10^{-4}$.
Despite marginal coupling, growth is still slower than dynamical and has $\sim$ AU scales for standard disk parameters.

We now extend the analysis to $\QDg = 1$ and $\taus = 1$, the center of our parameter space. Solution of the full dispersion relation shows that density waves are damped and the fastest growing neutral mode has $\gamma \approx 0.1$ and $\mk \approx 0.2$.  We make use of this result in comparing to simulations below.

\section{Discussion}\label{sec:disc}
\subsection{Comparison to Simulations}\label{sec:sims}
We now compare our linear instability with simulations that produce gravitational collapse of $\taus \sim 1$ solids \citep{nature07,jym09}.  We show that the non-linear gas dynamics included in the simulations gives faster growth than secular GI predicts.
  The relevant linear growth rates for $\taus = 1.0$ are given in \S\ref{sec:marg}.  The fact that the simulations include solids with somewhat tighter coupling (smaller $\taus$) enhances the differences highlighted here.

 \citet[hereafter Jetal07]{nature07} find gravitational collapse occurs very rapidly, within a few orbits of introducing self-gravity.  These simulations have turbulence generated by the MRI and by particle-gas interactions: both SI (streaming instabilities, \citealp{YG05}) and vertical shear instabilities.  Turbulent fluctuations in Jetal07 have a Mach number $\ve/c\gs \approx 0.05$, which we characterize as $\alpha \approx 0.05^2 = 2 \times10^{-3}$.  Their gas disk is reasonably massive with $Q\gs \sim 15$.  The collapse of solids occurs not from a smooth background but from a ring-like clump  generated by trapping in pressure maxima and SI-clumping.  Fig.\ 2 of Jetal07 shows that this ring has a width of $\sim 0.2 h\gs$.  The mean particle surface density is roughly a factor of 10 higher than the global average, giving $Z \sim 0.1$.  This value is a generous assessment of the mean surface  density of the knotted ring.  However, we now show that it is not generous enough to give agreement with linear theory.
 
The linear growth rates presented above give collapse in a few orbital times when $\QDg \sim 1$.  This measure of self-gravity in turn requires $Z \sim 0.5$, from \Eq{eq:QDg} and the values of $\alpha$ and $Q\gs$ inferred above.  This $Z$ value is larger than the average value of the simulation clumps.  Making the discrepancy worse, the linear theory requires an enhanced $Z$ over an extended radial distance. The relevant dimensionless wavenumber $\mk \sim 0.2$ corresponds to
\begin{equation} 
{\lambda \over h\gs} \sim {\lambda_G \over (0.2) h\gs} \sim {10 \pi \sqrt{\alpha} \over \QDg} \sim 1\, ,
\end{equation} 
\ie a wavelength that equals the gas scale height, wider than the clumps in the simulations.  Combining the mean surface density and  width discrepancies gives an even more significant discrepancy in the total mass required for rapid collapse.

This comparison shows that the collapse in simulations is faster and on smaller scales than in linear theory.   A deeper understanding of how SI clumping interacts with self-gravity is needed.  One factor is that SI (and perhaps other) clumping can reduce turbulent diffusion.   \citet{jym09} show that turbulent stirring of particles is weaker than average in dense clumps.  Less radial diffusion (lower effective $\alpha$) will enhance secular GI and give better agreement with detailed simulations.

\subsection{Planetesimal Sizes and Solar System ``Clans"}\label{sec:SS}
A major goal of planetesimal formation models is to explain the sizes of planetesimals.  This is useful as input for later stage models of coagulation \citep{gls04, bk06, kb06} and debris disks \citep{kb10}, which in turn evaluate the viability of planetesimal formation models.

Minor planets in the Solar System offer by far the most detailed constraints.  \citet{bornbig} claim that the largest asteroids with diameters $D \gtrsim 100$ km were ``born big" by a collapse process.  They argue that coagulation and collisional evolution models alone cannot reproduce the size distribution of the largest asteroids together with surface and shock age conditions.   \citet{stu10}, however, argues that coagulation of small, $D \sim 0.1$ km, planetesimals (with undetermined origins) could give better agreement.

In the outer Solar System, \citet{nyr10} propose that similar mass $D \gtrsim 50$ km binary Kuiper Belt objects could form by late stage fission of a collapsing proto-planetesimal.   The matching colors of binary companions \citep{ben09} supports this hypothesis.

Collapsing planetesimals with equivalent $D \gtrsim 100$ km form in  the simulations described in \S\ref{sec:sims}.  Since we do not model the non-axisymmetric fragmentation of secularly unstable rings, we do not make specific predictions of planetesimal sizes for comparison.  However $D \gg 1$ km is expected since the surface density when the ring fragments will greatly exceed the original $\Sigma$.

We do predict that each secularly collapsing ring will produce a clan\footnote{``Clan" purposely avoids ``family,"  which already has a specific meaning.}  of fairly identical planetesimals.  The solids in a ring will be well mixed during the slow initial collapse.   Since a disk's chemical composition and mass density should vary smoothly with distance and time (and further will have sharp transitions at ice lines), each ring inherits unique conditions during its birth.
Neighboring rings, or a subsequent generation, produce a planetesimal clan with different chemical composition and possibly sizes.  

The radial zonation of spectrally distinct asteroid classes \citep{gted82} supports this fragmenting ring hypothesis, and is difficult to explain with gradual collisional evolution.  The Kuiper Belt contains dynamical classes with different colors  \citep{dor08} and size distributions \citep{bern04}.  These signs may merely point to dynamical mixing of a continuously variable formation environment.  More detailed study and evolutionary modeling will help clarify if populations are distinct.  This could strengthen evidence for clans of planetesimals from the same birth ring.

\subsection{Radial Drift and Metallicity Thresholds}\label{sec:growthdisc}
Radial drift is usually the dominant constraint on secular GI growth times, but it might be over-restrictive.  Growth over several drift times, $\tdr$, is possible because $\tdr$ increases as the orbital radius shrinks [\Eq{eq:tdrifttight}].  Moreover, since drift pile-ups increase the surface density of solids \citep{ys02,yc04}, collapse could accelerate over several $\tdr$.  While we cannot include this sort of global evolution in our local analysis, we use drift pileups (and other enrichment mechanisms, see \citealp{cy10}) to motivate consideration of super-Solar $Z$ values.

Drift times would be much longer in special regions of the disk where the radial pressure gradient (and $\eta$) decreases.  Pressure maxima halt radial drift  and collect solids \citep{whi72}.  Enhancing $Z$  in turn accelerates secular GI.  Moreover the Keplerian gas motion in a pressure maximum has no headwind to stir particles.
Thus pressure maxima are a potential goldmine for planetesimal formation. 

Unfortunately the existence and properties of pressure maxima remain uncertain.  And appealing to special locations must also explain why planetesimal formation appears to have occured throughout the Solar System (and, with less detailed knowledge, throughout exoplanet and debris disk systems).  A possible downside to pressure maxima is that they might be associated with turbulence, as with MRI induced zonal flows  \citep{jyk09} or hydrodynamic vortices \citep{lith09}.  Whether turbulent stirring offsets the advantages of pressure maxima is uncertain and may depend on particle size. 

We now consider  turbulence driven by particle-gas interactions (for the standard case that radial pressure gradients do not vanish).     Both vertical shearing \citep{gw73, cdc93} and streaming \citep{YG05,yj07,jy07} instabilities can drive midplane mixing.   \citet{stu95} famously argued that vertical shear instabilities would prevent standard GI.  We show that  secular GI is not impeded for sufficiently super-Solar $Z$.

Particle-gas interactions stir the particle layer to a thickness $h \sim \eta R$, a value supported by analytic theory  \citep{stu95,  sek98, ys02} and simulation \citep{jym09, lee10}. To achieve this thickness, turbulence adjusts to a level
\begin{equation} \label{eq:alphaeta}
\alpha_\eta \approx \taus (\eta R/h\gs)^2 \approx 5 \times10^{-7} {a_{\rm mm} f_T \over \Fg m_{\ast}} \au^2
\end{equation} 
using \Eq{eq:hstd}.   For secular GI to proceed we require  $\alpha_\eta < \alpha_{\rm max}$ from   \Eq{eq:alphadrift}, which gives
\begin{equation} 
Z  > Z_\eta \approx {\eta R \over h\gs} \left(2\eta Q\gs \right)^{1/4} \approx 0.065 {f_T \over \sqrt{\Fg m_\ast}} \au^{1/4} \, , 
\end{equation} 
assuming $\rho > \rho\gs$ in \Eqs{eq:rhopg}{eq:fdr}.  This threshold requires substantial enrichment, especially if gas depletion lowers $\Fg$.  It provides our only reason to favor secular GI in more massive disks.

However, particle driven turbulence may weaken at lower $Z$. 
\citet{ys02} predicted a threshold to suppress vertical stirring of $\taus \ll 1$ solids at
\begin{equation} \label{eq:YS}
Z > Z_{\rm YS} \approx {\eta R \over h\gs} \approx 0.045 \sqrt{f_T \over m_\ast}\au^{1/4} \, .
\end{equation} 
\citet{jym09} found a threshold for strong SI clumping of $Z \gtrsim 0.015 = Z_{\rm YS}/3$ with particle sizes $0.1 \leq \taus \leq 0.25$ in 3D, vertically stratified simulations. 

\citet{bs10a} both confirmed the SI-clumping threshold and showed that it goes away (or shifts to higher $Z$ as they ``only" went up to $Z = 0.03$) when smaller sizes are included.   \citet{bs10b} show that the clumping threshold varies roughly linearly with the strength of  radial pressure gradients, as in \Eq{eq:YS}.

The existence of metallicity thresholds offers hope that $\taus \ll 1$ solids can overcome midplane stirring and undergo GI.  More simulations are warranted, but the direct simulation of $\taus \ll 1$ solids is computationally intensive, especially in large domains and over the long time scales needed for secular GI.

\subsection{Validity of the Model Equations}\label{sec:valid}
We now discuss the validity of our linearized equations of motion [eq.\ (\ref{eq:eom})] as a model of midplane particle dynamics.  

Radial drift does not appear in (or arise from) our momentum equations because a Galilean transformation removes steady drift and the locally constant radial pressure gradient.  Basic disk quantities ($\varSigma$, $\varOmega$, $\taus$) will change on the radial drift time scale, $\tdr$.  We avoid this complication by  conservatively restricting our attention to modes that grow faster than $\tdr$. 
Drift speeds also cause a transition to non-linear, turbulent drag laws, addressed in appendix \ref{sec:dragapp}.

By using a single value of the stopping time, we do not consider particle size distributions.  Since growth rates and wavelengths vary with $\taus$, it is worth considering whether a realistic dispersion of sizes could collapse coherently.  \citet{war00} demonstrated that a ``bimodal" disk with two particle sizes is subject to secular GI.  For a given $\taus$, a range of wavelengths around the fastest growing mode are unstable [see \Eq{eq:dispI} and \Fig{fig:nm_vs_dw}].  Thus a range of particle sizes can compromise and collapse at a common wavelength.  Depending on the width of the size distribution, the growth rate will decline relative to the mono-dispersive case.  

Using fluid equations to model the dynamics of solids subject to gas drag was discussed by   \citet{YG05}.  The issue of the appropriate particle effective pressure and viscous forces to include for this type of particle fluid has not been rigorously explored.  For this paper we use a standard relation between random velocity and pressure [the second term on the RHS of \Eq{eq:xmom}].  Since mass diffusion is a stronger stabilizing influence, the form of the effective pressure should not be too significant.

We neglect viscous forces, \ie terms of the form $D \p^2 u/\p x^2$ and $D \p^2 v/\p x^2$ in \Eqs{eq:xmom}{eq:ymom}, respectively.   It is possible for such viscous forces to introduce instabilities, as \citet{st95} show for planetary rings.   For our system however, drag forces dominate viscous forces. Thus we conclude that turbulence is more effective at diffusing solids than viscously transferring their momentum.  Furthermore, the precise form of viscous forces is uncertain, so we drop them for simplicity. 

We address the neglect of detailed gas dynamics in the following section.

\subsection{Previous (\& Future) Work on Secular GI}\label{sec:prev}
\citet{gw73} noted that: ``The frictional effect of gas drag does destabilize axisymmetric perturbations for wavelengths larger than [$\lambda_G$]."  Thus they were well aware of the existence of secular GI, but did not include it in their calculations.  \citet{war76,war00} derived and analyzed secular GI that is similar to this work, but neglected the mass diffusion induced by gas turbulence.  Indeed we can reproduce the dispersion relations in equation (A-22) of \citet{war76} and equivalently equation (31) of \citet{war00} by setting $\QDp = 0$ in our \Eq{eq:disp}. \citet{y05a,y05b} analyzed a similar dispersion relation, adding a more sophisticated model for turbulent stirring, but still neglecting turbulent radial diffusion. 

These (and our) analyses model the dynamics of  a single ``fluid," the particles.  Gas motions are in steady state.  A two fluid analysis evolves the aerodynamically coupled motion of solids and gas.   \citet{spi72} considered the two fluid Jeans problem, i.e. self-gravity in a static, non-rotating, infinite medium.  He showed that solids collapse relative to a static gas background on scales below the Jeans length.  The Jeans length in a gravitationally stable gas disk is quite large
 \begin{equation} 
\lambda_{\rm J} = c\gs \sqrt{\pi \over G \rho\gs} \approx 0.9 {f_T^{3/4} \au^{9/8} \over \sqrt{\Fg} m_\ast^{1/4}}\, {\rm AU}
\end{equation} 
supporting our assumption that long wavelength secular GI behaves in the single fluid limit.  However the role of disk dynamics remains unclear.

\citet{cor81} considered two fluid GI  in a thin disk.  They too conclude that solids do collapse through a stationary gas over a broad range of wavelengths.  A reanalysis of their model, including the effects of turbulent diffusion, would help clarify the upper wavelength cutoff to single fluid behavior and further explore two fluid phenomena.

The two fluid GI model of \citet{nvc91} numerically evaluates the linear growth of global non-axisymmetric modes, whose behavior depends on reflection at disk boundaries.  The relevance of these modes for planetesimal formation is unclear at this time, but non-axisymmetric modes should not be forgotten.

In the absence of self-gravity the two fluid model produces streaming instabilities, \ie SI \citep{YG05}.  Vertical speeds are required, thus height-integrated two fluid models like \citet{cor81} do not capture SI.   The interaction of secular GI with the non-gravitational SI clumping would thus require a 3D (or 2.5D for axisymmetric motion) two fluid model.  The SI radial wavelength, $\lambda_{\rm SI} \approx \taus \eta R$, does exceed the standard GI wavelength as
\begin{equation} 
{\lambda_{\rm SI} \over \lambda_G} \approx 30 \taus {f_T \over \Fg Z_\%}\, ,
\end{equation} 
provided $\taus$ is not very small.   For the longer wavelengths of secular GI, the interaction with smaller scale SI clumping is less clear.  

 \citet{gp00} studied secular GI and pure drag instabilities in a (single) single fluid model.  Their drag instability produces clumping and is analogous to SI (see CY10).  Drag instabilities can occur in their single fluid height-integrated model because they model drag as a surface density-dependent turbulent stress on the midplane layer.  As their formulation differs from a stopping time based approach, a direct comparison is difficult.  Goodman \& Pindor include viscous forces (that we ignore),  offering evidence that they do not suppress secular GI.
 
\citet{sek83} considers a different type of single fluid GI, that of a  midplane layer with perfectly coupled gas and ``dust" ($\taus \lll 1$ solids).  This model complements our treatment of collapse through the gas.  Sekiya finds a buckling mode where the dusty gas incompressibly rises from the midplane when the total density exceeds $\sim \rho_{\rm R}$ [\Eq{eq:rhoR}].  The sedimentation of dust from this blob could then give planetesimals.  However turbulent diffusion (which was not considered) should still be an obstacle to collapse.  Thus the relevance of $\rho_{\rm R}$ for small $\taus$ remains unclear.  A thorough investigation of the two fluid problem is needed to relate these different limiting cases.

\section{Conclusion}\label{sec:conc}
We explore gravitational instability (GI) acting on the particle sublayer in protoplanetary disks.  Gas drag introduces a class of secular (\ie dissipative) GI that differs from standard (dissipation free) GI.   We extend previous work by including mass diffusion due to turbulence.  We focus mainly on the behavior of small solids, with stopping times $\taus \ll 1$, to study their possible gravitational collapse into planetesimals.   We also consider $\taus \gtrsim 1$ to understand the transition from secular to standard GI.

We prove that secular GI always produces collapsing modes.  This feature qualitatively differs from standard GI, which must satisly threshold conditions on Toomre's $\QT \lesssim 1$, or equivalently a critical, Roche-like density.   
While secular GI is formally ever-present, it can only produce planetesimals if collapse is more rapid than particle radial drift and disk dispersal.   The growth rates and wavelengths of secular GI depend sensitively on properties of the gas and particle disks, which we briefly summarize.

Gas drag is a double-edged sword.  It introduces the possibility of secular GI, but gives progressively slower growth for smaller $\taus \ll 1$  by limiting infall to terminal speeds.  Thus with other things (turbulence and $Z$) being equal, collapse is favored for larger solids and in the outer disk where gas densities are lower.

Enhancing the disk ``metallicity" $Z$ promotes GI, both by increasing the strength of particle self-gravity and by slowing radial drift.  A higher total (gas \& particle) disk mass does not directly benefit secular GI, because higher gas densities exert stronger drag (lower $\taus$). 

Turbulence slows growth, primarily by diffusively smoothing small scale collapse.   Slow collapse occus over many orbits and thus takes the form of sheared-out rings that contract radially.  These initially wide rings contain up to $\sim 0.1 M_\oplus$ of solids, which will eventually fragment into lower mass planetesimals by a process not described here. 

Whether small solids can form planetesimals by secular GI depends on what constitutes a ``realistic" level of midplane turbulent diffusion, which we characterize as an $\alpha$ parameter.  Our upper limits on $\alpha$ are low, $\sim 10^{-8}$ --- $10^{-3}$, depending on disk parameters, as shown in \Fig{fig:almax}.
 
Our diffusive $\alpha$ can be much smaller than the more commonly used ``viscous" $\alpha_\nu$ (subscript added to differentiate) that measures accretion stresses and angular momentum transport  \citep{pri81}.    Disk accretion may occur in active surface layers, with the midplane a quiescent dead zone where the MRI (magneto-rotational instability) does not operate \citep{gam96,pbc11}.
Furthermore, angular momentum transport could be dominated by magnetic Maxwell stresses or waves that do not contribute to turbulent diffusion.   This argument applies (e.g.) to waves launched in active layers that penetrate into the dead zone \citep{fs03,omm07}.  Finally the tendency of particles to concentrate in turbulence can counteract diffusion and reduce the effective $\alpha$ (see \S\ref{sec:sims}, CY10).  We conclude that disk accretion, with observationally constrained values of $10^{-4} \lesssim \alpha_\nu \lesssim 0.1$ \citep{and10}, does not prevent planetesimal formation with significantly lower values of diffusive $\alpha$.

The strictest barrier to small particle GI is still posed by particle-driven turbulence, as proposed by \citet{stu95}.  Overcoming this obstacle requires localized reductions in radial pressure gradients and/or enhancing  $Z$ to super-Solar values as discussed in \S\ref{sec:growthdisc}.  The strong correlation of exoplanets with host star metallicity \citep{john10} almost requires that planet formation have a strong $Z$ dependence. 

Further progress can be made by better understanding the relation between secular GI and non-gravitational particle concentration by streaming instabilities and turbulent fluctuations.   Dynamical theory and simulation complements interdisciplinary research on planet formation that spans collisional physics and experiments, small bodies in the Solar System, and astronomical studies of disks and exoplanets.

\acknowledgements
I thank Jeremy Goodman for suggesting Descartes' rule of signs to prove instability and an anonymous referee of a previous submission for suggesting the importance of radial turbulent diffusion.  Comments from Kaitlin Kratter, Scott Kenyon and Xeuning Bai greatly improved the presentation.  Portions of this project were supported by the {\it NASA} {\it Astrophysics Theory Program} and  {\it Origins of Solar Systems Program}  through grant NNX10AF35G.\\

\if\bibinc n
\bibliography{refs}
\fi

\if\bibinc y

\fi
 \clearpage

\appendix
\section{Nonlinear Drag Laws}\label{sec:dragapp}
Here we provide complete prescriptions to calculate the dimensionless stopping time $\taus \equiv \varOmega \ts$, that gives a drag acceleration $\vc{f}_{\rm drag} = - \Delta \vc{V}/\ts$ on a particle of mass $m = (4 \pi/3)\rho_\bullet a^3$ moving at a relative velocity $\Delta \vc{V}$.  We focus on turbulent drag forces which have a nonlinear dependence on relative velocity, as this case was not described in \S\ref{sec:drag}.  We also discuss the implications of non-linear drag for linear stability analyses such as ours.

For compact notation (following \citealp{ahn76}) we use the Knudsen number $K = \ell\gs/a$ and the Mach number $M = \Delta V/c\gs$.  For the relative speed, we use the equilibrium drift speed between solids and gas, 
\begin{equation} 
\Delta V =  \eta \varOmega R {\sqrt{4 \taus^2 + \taus^4} \over 1 + \taus^2}\, ,
\end{equation} 
\ie the vector sum of azimuthal headwind and radial drift, because it is much faster than linear perturbation speeds.  To include particle inertia, we could make the substitution (\citealp{nsh86}, see also equations [14] and [15] of \citealp{YG05})
\begin{equation} \label{eq:tausinert}
 \taus \rightarrow \taus/(1+\rho / \rho\gs)\, .
\end{equation} 
This complication --- making $\taus$ depend on particle density, $\rho$, and thus turbulence and settling --- is not justified for the current study.   The correction vanishes for large $\taus \gg 1 + \rho/\rho\gs$, and thus has minimal effect (since large $\taus$ is required in practice for turbulent drag). 

We assign $\taus$ as follows:
\begin{equation}\label{eq:taurules}
\taus = \left\{\begin{array}{ccc}
\tau_{\rm Ep}&\mathrm{if}& \frac{4}{9} < K \\
\tau_{\rm Sto}  \equiv {4 \over 9 K} \tau_{\rm Ep} &\mathrm{if}&\left( \frac{4}{3} \right)^5 M < K<  \frac{4}{9}\\
\tau_{\rm int} \equiv {9 \over 16} \left({K \over M}\right)^{2/5}  \tau_{\rm Sto}  &\mathrm{if}& \frac{M}{200} < K< \left( \frac{4}{3} \right)^5 M \\
\tau_{\rm turb} \equiv {200^{3/5} \over 4 M}\tau_{\rm Ep} &\mathrm{if}& K<\frac{M}{200} 
\end{array}\right.\, .
\end{equation}
Because the transitions themselves depend on $\taus$ via $M$, \Eq{eq:taurules} is best solved numerically.   We use integer coefficients to assure perfect matching at the transitions.  The prescription for turbulent drag is consistent with the standard turbulent drag coefficient $C_D \approx 0.44$, since we can identify
\begin{equation} 
C_D \equiv {2 m |\vc{f}_{\rm drag}| \over \pi a^2 \rho\gs \Delta V^2} = {8 \rho_\bullet a \over 3 \ts \rho\gs \Delta V} = {8 \tau_{\rm Ep} \over 3 \tau_{\rm turb} M} = {32 \over 3 \cdot200^{3/5}} \approx 0.44
\end{equation} 
Intermediate drag, given by $\tau_{\rm int}$, describes the transition from viscous drag to fully developed turbulent wakes. 

We now consider corrections to the drag force in linear perturbation analyses when turbulent drag applies.  In \Eq{eq:eom} we used a drag acceleration of the form
\begin{equation} \label{eq:dragpar}
\vc{f}'_{\rm drag} = -{\vc{v} \over \ts} 
\end{equation} 
where $\vc{v} = u \hat{x} + v \hat{y}$ is the linear perturbation velocity.  While the total drag acceleration is always (anti-) parallel to the drift speed, for quadratic drag the perturbed drag force is not exactly  (anti-) parallel to the perturbed velocity.   Perturbed drag forces are twice as strong along the direction of the background drift. 

To modify \Eq{eq:dragpar} for quadratic drag, start with the \emph{total} drag force
\begin{equation} 
\vc{f}_{\rm drag} = -{1 \over \ls}\left| \Delta\vc{V} - \vc{v}\right|\left( \Delta\vc{V} - \vc{v}\right)
\end{equation} 
The stopping length $\ls = |\Delta \vc{V}|\ts$ is velocity-independent.  The orientation of pressure gradient drift is
\begin{equation} 
\tan(\theta) \equiv {\Delta V_y \over \Delta V_x} = - {\taus \over 2}\, ,
\end{equation} 
and \Eq{eq:tausinert} would again give the inertial correction.  The steady drag force is just $-|\Delta \vc{V}|\Delta \vc{V}/\ls$.  The linear perturbation is
\begin{equation} 
\vc{f}'_{\rm drag} = - {1 \over \ts}  \left[\vc{v} + \left(u   \cos^2\theta + v { \sin2 \theta\over 2}  \right) \hat{x} +  \left(v  \sin^2\theta + u { \sin2 \theta\over 2} \right) \hat{y}\right]
\end{equation} 
 We do not include this correction, as it would mainly apply to large $\taus$ in the inner disk, not our primary focus.  Generalizations to intermediate (not quadratic) drag and to include vertical motion are possible.

\section{Regions IV \& V: Too Dynamically Cold}\label{sec:regapp}
We now consider two regions of our parameter space, which are unlikely to be of much astrophysical relevance.  We include their description for completeness and to better understand the limitations of the mathematical model.

\subsection{Region V: $\QDg^{3/2} \ll \taus \ll 1$} \label{sec:regV}
Diffusion is exceedingly weak in region V.  The condition $\QDg \ll \taus^{3/2}$ corresponds to
\begin{equation} \label{eq:alphaV}
\alpha \ll \alpha_{\rm V} \equiv \taus^3 Z^2/Q\gs^2 \approx 10^{-18} {a_{\rm mm}^3 Z_\%^2 \over \Fg f_T m_\ast}\au^5\, .
\end{equation} 
For comparison, this $\alpha$ is even lower than molecular viscosity, $\alpha _{\rm mv} \approx \ell\gs/h\gs \approx 10^{-12} \au^{3/2}/\Fg$, except in the outer disk.  In \Fig{fig:al6810}, the upturn in growth rates of the $\alpha = 10^{-10}$ curve beyond 30 AU is due to the transition to region V.

In Region V free-fall gravitational collapse is countered my mass diffusion.  Even though $\taus \ll1$,  collapse is so rapid that inertia exceeds drag forces.    Balancing free-fall collapse [\Eq{eq:ff}] with the diffusive time scale $t_{\rm diff} \sim \lambda^2/D\ps$, gives the wavelength, $\lambda_{\rm V} \sim D^{2/3}/(G \varSigma)^{1/3}$, and rate, $s_{\rm V} \sim (G \varSigma)^{2/3}/D^{1/3}$, or growth.  An expansion of the dispersion relation confirms that  $\gamma_{\rm V} = 1/\QDg^{2/3}$ and $\mk = 1/\QDg^{4/3}$.

With $\gamma_{\rm V} \gg 1/\taus$, collapse is even faster than the (short in orbital terms) stopping time.  This extremely rapid collapse introduces a physical inconsistency.  The standard assumption that $\taus \ll 1$ solids diffuse at the gaseous rate, $D = D\gs$, likely breaks down.  

\subsection{Region IV: $\QDg^{2} \ll 1/\taus \ll 1$} \label{sec:regIV}
Region IV is the large $\taus$ partner of region V.  The nominal requirement on $\alpha$ is again quite small.  The more fundamental concern is that stirring by self-gravity would prevent a disk of $\taus \gg 1$ solids from being so quiescent.  This was discussed in \S\ref{sec:regIII}.

Putting these concerns aside, the behavior in region IV is identical to region V: gravitational collapse balances diffusion.  This balance differs from region III, where the longer wavelength collapse is primarily opposed by pressure, not diffusion.
The only change from the region V derivation is that $D \sim D\gs/\taus^2$ for $\taus \gg 1$.  This gives $\gamma_{\rm IV} \sim (\taus/\QDg)^{2/3}$ and $\mk_{\rm IV} \sim (\taus/\QDg)^{2/3}$.  The collapse rate is much faster than dynamical with $\gamma_{\rm IV}\gg \taus \gg 1$, but this behavior does not appear in our disk models.


%

\end{document}